\documentclass[aps,a4paper,pre,amsmath,amsfonts,amssymb,nofootinbib]{revtex4}

\usepackage{graphicx}
\usepackage{epsfig}
\def\de{{\rm d}}
\def\s{\sigma}
\def\us{\underline{\sigma}}
\def\uJ{\underline{J}}
\def\t{\tau}
\def\ut{\underline{\tau}}

\def\eqd{\overset{\rm d}{=}}
\def\ed{\overset{\rm d}{=}}
\def\E{\mathbb{E}}
\def\I{\mathbb{I}}

\def\e{{\sf E}}
\def\te{\widetilde{\sf E}}
\def\A{{\cal A}}

\def\P{{\cal P}}
\def\Q{{\cal Q}}

\def\Z{{\cal Z}}
\def\oh{\overline{h}}
\def\ou{\overline{u}}

\def\cX{{\cal X}}

\def\oP{\overline{P}}
\def\oQ{\overline{Q}}
\def\tP{\widetilde{P}}
\def\tQ{\widetilde{Q}}

\def\hQ{\widehat{Q}}
\def\hP{\widehat{P}}

\def\halpha{\widehat{\alpha}}
\def\tSigma{\widetilde{\Sigma}}
\def\hx{\widehat{x}}
\def\da{{\partial a}}
\def\di{{\partial i}}
\def\reals{{\mathbb R}}
\def\cA{{\cal A}}
\def\integers{{\mathbb Z}}
\def\cN{{\cal N}}

\begin{document}
\date{\today}

\title{Clusters of solutions and replica symmetry breaking in\\
random $k$-satisfiability}

\author{Andrea Montanari}
\affiliation{Depts of Electrical Engineering and Statistics, Stanford
  University, USA.}
\author{Federico Ricci-Tersenghi}
\affiliation{Dipartimento di Fisica and INFM-CNR, Universit\`a di Roma
  La Sapienza, P. A. Moro 2, 00185 Roma, Italy.}
\author{Guilhem Semerjian}
\affiliation{LPTENS, Unit\'e Mixte de Recherche (UMR 8549) du CNRS et
  de l'ENS, associ\'ee \`a l'UPMC Univ Paris 06, 24 Rue Lhomond, 75231
  Paris Cedex 05, France.}

\begin{abstract}
  We study the set of solutions of random $k$-satisfiability formulae
  through the cavity method. It is known that, for an interval of the
  clause-to-variables ratio, this decomposes into an exponential
  number of pure states (clusters).  We refine substantially this
  picture by: $(i)$ determining the precise location of the clustering
  transition; $(ii)$ uncovering a second `condensation' phase
  transition in the structure of the solution set for $k \ge 4$.
  These results both follow from computing the large deviation rate of
  the internal entropy of pure states. From a technical point of view
  our main contributions are a simplified version of the cavity
  formalism for special values of the Parisi replica symmetry breaking
  parameter $m$ (in particular for $m=1$ via a correspondence with the
  tree reconstruction problem) and new large-$k$ expansions.
\end{abstract}

\maketitle

\section{Introduction}

An instance of $k$-satisfiability ($k$-SAT) consists in a Boolean
formula in conjunctive normal form whereby each elementary clause is the
disjunction of $k$ literals (a Boolean variable or its negation).
Solving it amounts to determining whether there exists an assignment
of the variables such that at least one literal in each clause
evaluates to true.  The $k$-SAT problem plays a central role in the
theory of computational complexity, being the first decision problem
proven to be NP-complete~\cite{NP} (for all $k \ge 3$). Its
optimization (minimize the number of unsatisfied clauses) and
enumeration (count the number of optimal assignments) versions are
defined straightforwardly and are also hard from the computational
point of view.

Random $k$-satisfiability is the ensemble defined by drawing a
uniformly random formula among all the ones involving $M$ $k$-clauses
over $N$ variables. Equivalently, each of the $M$ clauses is drawn
uniformly over the $2^k \binom{N}{k}$ possible ones, independently
from the others.  It was observed empirically earlier
on~\cite{transition_exp} that, by tuning the clause density $\alpha =
M/N$, this ensemble could produce formulae which were hard for known
algorithms. Hardness was argued to be related to a sharp threshold in
the satisfiability probability, emerging as $N\to\infty$ with $\alpha$
fixed.  More precisely, it is believed that there exists a constant
$\alpha_{\rm s}(k)$ such that random formulae are with high
probability\footnote{Here and below `with high probability' (w.h.p.)
  means with probability converging to $1$ as $N\to\infty$.}
satisfiable if $\alpha < \alpha_{\rm s}(k)$ and unsatisfiable if
$\alpha > \alpha_{\rm s}(k)$.  The existence of a sharp threshold was
proven in \cite{Friedgut}, with, however, a critical point
$\alpha_{\rm s}(k,N)$ which might not converge when $N \to \infty$.
Despite important progresses \cite{review_lb,review_ub,largek} the
rigorous proof of the existence and determination of $\alpha_{\rm
  s}(k)$ remains a major open problem (with the notable exception of
$k=2$ \cite{k2}).

The connection between threshold phenomena and phase transitions
spurred a considerable amount of work
\cite{MoZe,BiMoWe,MeZe,MePaZe,MeMeZe} using techniques from the theory
of mean field spin glasses \cite{Beyond}.  The main outcomes of this
approach have been: $(i)$ A precise conjecture on the location of the
satisfiability threshold $\alpha_{\rm s}(k)$~\cite{MeZe,MeMeZe};
$(ii)$ The suggestion~\cite{BiMoWe,MeZe} for $k \ge 3$ of another
transition at $\alpha_{\rm d}(k) < \alpha_{\rm s}(k)$ affecting the
geometry of the solutions space; $(iii)$ Most strikingly, the proposal
of a new and extremely effective message passing algorithm, Survey
Propagation (SP)~\cite{MeZe,MePaZe}. This exploits a detailed
statistical picture of the solution space to efficiently find
solutions.

According to statistical physics studies, in the intermediate regime
$\alpha \in [\alpha_{\rm d}(k),\alpha_{\rm s}(k) ]$ solutions tend to
group themselves in clusters that are somehow disconnected.  As
$\alpha$ increases, the number of these clusters decreases.  The
satisfiability transition is thus due to the vanishing of the number
of clusters, which still contain a large number of solutions just
before $\alpha_{\rm s}(k)$.  The phase transition at $\alpha_{\rm
  d}(k)$ has been referred to as ``clustering phase transition'' or
``dynamic phase transition'' depending on the feature emphasized. Its
nature and location, as well as a refined description of the regime
$\alpha \in [\alpha_{\rm d}(k),\alpha_{\rm s}(k) ]$ will be the main
topic of this paper.  More precisely:
\begin{enumerate}
\item[$(i)$] We will argue that previous determinations of
  $\alpha_{\rm d}(k)$ \cite{MeZe,MePaZe,MeMeZe} have to be corrected when
  fluctuations of the cluster sizes are taken into account;
\item[$(ii)$] We will uncover (for $k \ge 4$) a new `condensation'
  phase transition at $\alpha_{\rm c}(k)\in [\alpha_{\rm
    d}(k),\alpha_{\rm s}(k)]$.  For $\alpha\in [\alpha_{\rm
    d}(k),\alpha_{\rm c}(k)]$ the relevant clusters are exponentially
  numerous.  For $\alpha\in [\alpha_{\rm c}(k),\alpha_{\rm s}(k)]$
  most of the solutions are contained in a number of clusters that
  remains bounded as $N\to\infty$.
\end{enumerate}

The paper is organized as follows.  In Section~\ref{sec_reminder} we
recall some general features of mean-field disordered models,
emphasizing the notions of dynamical transitions and replica symmetry
breaking.  In Section~\ref{sec_cavity} we define more precisely the
ensemble of random formulas studied and describe the replica symmetric
(RS) and one step of replica symmetry breaking (1RSB) approach to this
model. We then apply the program of Section~\ref{sec_reminder} to the
random $k$-satisfiability problem and present our main results in
Section~\ref{sec_transitions}. For the sake of clarity some
technicalities of the 1RSB treatment are presented shortly afterward,
see Section~\ref{sec_1rsb_tricks}. To complement these results, which
are partly based on a numerical resolution of integral equations, we
present in Section~\ref{sec_large_k} an asymptotic expansion in the
large $k$ limit which gives further credit to our theses.  We draw our
conclusions in Sec.~\ref{sec_conclusions}. Technical details are
deferred to three appendices.

A short account of our results has been published in~\cite{letter},
and a detailed analysis of the related $q$-coloring problem
in~\cite{FlLe}. While the present work was being finished two very
interesting papers confirmed the generality of the results
of~\cite{letter}.  The first concerned 3-SAT~\cite{k3_Zhou} and the
second bi-coloring of random hypergraphs~\cite{Ramenza}.

\section{Mean-field disordered systems}
\label{sec_reminder}

The goal of this section is to provide a quick overview of the cavity
method \cite{Beyond,Talagrand_book}. We will further propose a more
precise mathematical formulation of several notions that are crucial
in the statistical physics approach.
%
%
\subsection{Statistical mechanics and graphical models}

Let us start by considering a general model defined by:

$(1)$ A factor graph~\cite{fgraphs}, i.e.\ a bipartite graph
$G=(V,F,E)$. Here $V$, $|V|=N$, are `variable nodes' corresponding to
variables, $F$, $|F|=M$, are `function (or factor) nodes' describing
interactions among these variables, and $E$ are edges between
variables and factors. Given $i\in V$ (resp.\ $a\in F$), we shall
denote by $\di = \{a\in F: (ia)\in E\}$ (resp.\  $\da = \{i\in V:
(ia)\in E\}$) its neighborhood.  Further, given $i,j\in V$, we let
$d(i,j)$ be their graph theoretic distance (the minimal number of
factor nodes encountered on a path between $i$ and $j$).

$(2)$ A space of configurations $\cX^V$, with $\cX$ a finite alphabet,
(a configuration will be denoted in the following as
$\us=(\sigma_1,\dots,\sigma_N) \in \cX^V$).  For any set $A\subseteq
V$, we let $\us_A = \{\sigma_i:\, i\in A\}$.

$(3)$ A set of non negative weights $\{w_a:\, a\in F\}$,
$w_a:\cX^{\da}\to \reals_+$, $\us_{\da}\mapsto w_a(\us_{\da})$. In the
case of \emph{constraint satisfaction problems}, these are often taken
to be indicator functions (more details on this particular case will
be given in Sec.~\ref{sec_reminder_csp}).

Given these ingredients, a measure over $\cX^V$ is defined as
\begin{eqnarray}
\mu_N(\us) = \frac{1}{Z_N}\, w_N(\us) \ , \qquad w_N(\us) =
\prod_{a\in F}w_a(\us_{\da})\, .\label{eq:Factor}
\end{eqnarray}
This is well defined only if there exists at least one configuration
$\us^*$ that makes all the weights strictly positive, namely
$w_a(\us^*_{\da})>0$ for each $a$.  We will assume this to be the case
throughout the paper (i.e.\ we focus on the `satisfiable' phase).
Further, it will be understood that we consider sequences of graphs
(and weights) of diverging size $N$ (although we shall often drop the
subscript $N$).

An important role is played by the large-$N$ behavior of the
partition function $Z_N$. This is described by the free-entropy
density\footnote{One usually assumes that the limit exists. If the
  model is disordered, almost sure limit can be used, or,
  equivalently, $\log Z_N$ is replaced by its expectation.}  ,
\begin{equation}
\phi = \lim_{N \to \infty} \frac{1}{N} \log Z_N \ , 
\qquad Z_N =\sum_{\us} w_N(\us) \ .
\end{equation}
%
%
\subsection{Pure states and replica symmetry breaking}
\label{sec_reminder_rsb}

The replica/cavity method allows to compute a hierarchy of
approximations to $\phi$. This is thought to yield the exact value of
$\phi$ itself in `mean field' models. The hierarchy is ordered
according to the so-called number of steps of replica symmetry
breaking (RSB).  At each level the calculation is based on some
hypotheses on the typical structure of $\mu$, a pivotal role being
played by the notion of \emph{pure state}. Since this concept is only
intuitively defined in the physics literature, we propose here two
mathematically precise definitions. In both cases a pure state is a
(sequence of) probability measures $\rho_N$ on $\cX^N$.

\begin{itemize}
\item[$\bullet$]
\underline{Definition of pure states through correlation decay}

We define the correlation function of $\rho_N$ as 
\begin{eqnarray}
C_N(r) = \sup_{A,B:\, d(A,B)\ge r}\sum_{\us_A,\us_B}
\left|\rho_N(\us_A,\us_B)-\rho_N(\us_A)\rho_N(\us_B)\right|\, ,
\end{eqnarray}
where the $\sup$ is taken over all subset of variable nodes
$A,B\subseteq V$ such that the distance between any pair of nodes
$(i,j) \in A \times B$ is greater than $r$.  Then $\rho_N$ is a pure
state if this correlation function decays at large $r$. Technically,
we let $C_{\infty}(r) = \lim\sup_{N\to\infty}C_N(r)$, and require
$C_{\infty}(r)\to 0$ as $r\to\infty$.

\item[$\bullet$]
\underline{Definition of pure states through conductance}

We let the $(\epsilon, \delta)$-conductance of $\rho_N$ be
\begin{eqnarray}
{\mathfrak F}_N(\epsilon,\delta) = 
\inf_{\cA \subset
  \cX^{N}}\left\{\frac{\rho_N(\partial_{\epsilon}\cA)}{\rho_{N}(\cA)(1-\rho_N(\cA))}:\,
  \delta \le\rho_N(\cA)\le 1-\delta\right\}\, . 
\end{eqnarray}
Here the $\inf$ is taken over all subsets of the configuration
space. Further, letting $D$ denote the Hamming distance in $\cX^N$, we
defined the boundary of $\cA$ as $\partial_{\epsilon}\cA = \{\us\in
\cX^N\setminus\cA\,|\; D(\us,\cA)\le N\epsilon\}$.  With these
definitions $\rho_N$ is pure if its conductance is bounded below by an
inverse polynomial in $N$ for all $\epsilon$ and $\delta$ (while
non-pure states have a conductance which typically decays
exponentially with $N$).
\end{itemize}

These two definitions mimic the well-known ones on $\integers^d$ in
terms of tail triviality and extremality~\cite{Georgii}. Further, the
second one is clearly related to the behavior of local Monte Carlo
Markov chain dynamics. A small conductance amounts to a bottleneck in
the distribution and hence to a large relaxation time.  While we
expect them to be equivalent for a large family of models, proving
this is a largely open problem. Moreover we should emphasize that the
heuristic cavity method followed in this paper never explicitly uses
either of these definitions.

The hypotheses implicit in the cavity method can be expressed in terms
of the \emph{pure states decomposition} of $\mu$.  This is a partition
of the configuration space (dependent on the graph and weights) such
that the measure $\mu$ constrained to each element of this partition
is a pure state.  More precisely, let us call $\{ \A_\gamma
\}_{\gamma}$ a partition of $\cX^N$, and define
\begin{equation}
Z_\gamma = \sum_{\us \in \A_\gamma} w(\us) \ , \qquad
W_\gamma = \frac{Z_\gamma}{Z} \ , \qquad 
\mu_\gamma(\us) = \frac{1}{Z_\gamma} w(\us) \I(\us \in \A_\gamma ) \ .
\end{equation}
Clearly $\mu$ can be written as the convex combination of the
$\mu_{\gamma}$ with coefficients $W_{\gamma}$.  This defines a pure
state decomposition if: $(i)$ each of the $\mu_{\gamma}$ is a pure
state in the sense given above, $(ii)$ this is the `finest' such
partition, in the sense that the $\mu_{\gamma}$ are no longer pure if
any subset of them is replaced by their union.

Statistical physics calculations suggest that a wide class of mean
field models is described by one of the following `universal
behaviors'.  The terminology used here is inherited from the
literature on mean field spin
glasses~\cite{Giorgio_Houches,Leticia_Houches}.
\begin{itemize}
\item[{\sf RS}] Most of the measure is contained in a single element
  of the partition, namely $W_{\rm max} = \max_{\gamma} W_{\gamma}\to
  1$ as $N\to\infty$ (\emph{replica symmetric}).
\item[{\sf d1RSB}] Most of the measure is carried by $\cN\doteq
  e^{N\Sigma_*}$ pure states\footnote{Here and in the following
    $\doteq$ means equality at the leading exponential order.}, each
  one with a weight $W_{\gamma}\doteq e^{-N\Sigma_*}$ (\emph{dynamical
    one-step replica symmetry breaking}).
\item[{\sf 1RSB}] The measure condensates on a subexponential number
  of pure states, namely, if $W_{[\gamma]}$ is the weight of the
  $\gamma$-th largest state, then $\lim_{n \to \infty} \lim_{N \to
    \infty} \sum_{\gamma=1}^n W_{[\gamma]} =1$ (\emph{one step replica
    symmetry breaking}).
\end{itemize}
The reader will notice that this list does not include full replica
symmetry breaking phases, in which pure states are organized according
to an ultrametric structure. While this behavior is as generic as the
previous ones, our understanding of it in sparse graph models is still
rather poor.

We are mostly concerned with families of models of the type defined in
Eq.~(\ref{eq:Factor}) indexed by a continuous parameter $\alpha$ (such
as the clause density in $k$-SAT).  In this setting, the above
behaviors often appear in sequence as listed above when the system
becomes more and more constrained (e.g.\  as $\alpha$ is increased in
$k$-SAT). The different regimes are then separated by phase
transitions: the `dynamical' or `clustering' phase transition from
{\sf RS} to {\sf d1RSB} (at $\alpha_{\rm d}$) and the `condensation'
phase transition between {\sf d1RSB} and {\sf 1RSB} (at $\alpha_{\rm
  c}$). The paradigmatic example of such transitions is the
fully-connected $p$-spin model~\cite{Giorgio_Houches,Leticia_Houches},
where they are encountered upon lowering the temperature.

Let us stress that the above definitions are insensitive to what
happens in a fraction of the space of configurations of vanishing
measure. For instance, we neglect metastable states whose overall
weight is exponentially small\footnote{In the fully connected models
  such metastable states are indeed seen as solutions of the
  Thouless-Anderson-Palmer equations, well above the dynamical phase
  transition.}.

A convenient tool for distinguishing these various behaviors is the
replicated free-entropy~\cite{Mo},
\begin{equation}
  \Phi(m) = \lim_{N \to \infty} \frac{1}{N} \E \log\left\{ 
    \sum_{\gamma} Z_\gamma^m \right\}\ ,
\label{eq_def_Phi}
\end{equation}
where $m$ is an arbitrary real number (known as Parisi replica symmetry 
breaking parameter)
which allows to weight differently the various pure states according to
their sizes.  Suppose indeed that the number of pure states $\gamma$
with internal free-entropy density $\phi_\gamma = (\log Z_\gamma)/N$
behave at leading order as $\exp\{ N \Sigma(\phi_\gamma)\}$, where
$\Sigma(\phi)$ is known as the complexity (or configurational entropy)
of the states. The sum in (\ref{eq_def_Phi}) can then be computed by
the Laplace method; if one assumes for simplicity that $\Sigma$ is
positive on an interval $[\phi_-,\phi_+]$, this leads to
\begin{equation}
\Phi(m) = \sup_{\phi \in [\phi_-,\phi_+]} [\Sigma(\phi) + m \phi ] \ .
\label{eq_Legendre}
\end{equation}
Provided $\Sigma$ is concave, it can be reconstructed in a parametric
way from $\Phi(m)$ by a Legendre inversion~\cite{Mo},
\begin{equation}
\Sigma(\phi_{\rm int}(m)) = \Phi(m) - m \Phi'(m) \ , \qquad
\phi_{\rm int}(m) = \Phi'(m) \ ,
\label{eq_Legendre_inversion}
\end{equation}
where $m$ is such that the supremum in (\ref{eq_Legendre}) lies in the
interior of $[\phi_-,\phi_+]$, which defines a range
$[m_-,m_+]$. Usually $\Sigma$ vanishes continuously at $\phi_+$. As
explained below, when zero energy states are concerned $\phi_{\rm
  int}(m)$ coincides with the internal entropy of such states.
Note that a given value of $m$ selects the point of the curve
$\Sigma(\phi)$ of slope $-m$; in particular the value $m=0$
corresponds to the maximum of the curve.

The replica/cavity method at the level of one step of replica symmetry
breaking allows to compute the replicated free-entropy $\Phi(m)$ under
an appropriate hypothesis on the organization of pure states. The
various regimes can be distinguished through the behavior of this
function, namely
\begin{itemize}
\item[{\sf RS}] $\Phi(m) = m\phi_*$, where $\phi_*$ is the
  contribution of the single dominant pure state, $Z_{[1]} \doteq
  e^{N\phi_*}$.
\item[{\sf d1RSB}] $\Phi(m)/m$ achieves its minimum for $m\in[0,1]$ at
  $m=1$, with $\Sigma_*=\Phi(1)-\Phi'(1)>0$. Then the measure $\mu$
  decomposes into approximately $e^{N\Sigma_*}$ pure states of
  internal free-entropy $\Phi'(1)$.
\item[{\sf 1RSB}]$\Phi(m)/m$ achieves its minimum over the interval
  $[0,1]$ at $m_{\rm s}\in(0,1)$. Then the ordered sequence of weights
  $W_{[1]}\ge W_{[2]}\ge W_{[3]}\ge \cdots$ keep fluctuating in the
  thermodynamic limit, and converges to a Poisson-Dirichlet
  process~\cite{Ruelle} of parameter $m_{\rm s}$.  The internal
  free-entropy of these states is $\Phi'(m_{\rm s})$.
\end{itemize}
In all these cases the total free-entropy density is estimated by
minimizing $\Phi(m)/m$ in the interval $[0,1]$.

%
%
\subsection{Cavity equations}
\label{sec_reminder_sparse}

We shall now recall the fundamental equations used within the 1RSB
cavity method and propose a somehow original derivation.  In the
following we will be interested in factor graphs that converge
locally\footnote{More precisely, any finite neighborhood of a
  uniformly chosen random vertex converges to a tree.}  to trees in
the thermodynamic limit.

In consequence, let us first consider the case of a model of type
(\ref{eq:Factor}) whose underlying factor graph is a tree, and discuss
later how the long loops are taken into account by the cavity method.
Tree factor graph models are easily solved by a `message passing'
procedure~\cite{fgraphs}.  One associates to each directed edge from
factor $a$ to variable $i$ (resp.\ from $i$ to $a$) a ``message''
$\eta_{a \to i}$ (resp.\ $\eta_{i \to a}$). Messages are probability
measures on $\cX$.  On trees, they can be defined as the marginal law
of $\sigma_i$ with respect to the modified factor graph $G_{a \to i}$
(resp.\ $G_{i \to a}$) where all factor nodes in $\di \setminus a$
(resp.\ the factor node $a$) have been removed.  Simple computations
yield the following local equations between messages,
\begin{eqnarray}
\eta_{a \to i} &=& f_{a \to i}(\{\eta_{j \to a} \}_{j \in \partial a 
\setminus i}  ) \ , \qquad f_{a\to i}(\{\eta_{j \to a} \}) (\sigma_i) =
\frac{1}{z_{a \to i}(\{\eta_{j \to a} \} )}
\sum_{\us_{\partial a \setminus i}} 
w_a(\us_{\partial a}) \prod_{j \in \partial a \setminus i} 
\eta_{j \to a}(\sigma_j) \ , 
\label{eq_bp_atoi} 
\\
\eta_{i \to a} &=& f_{i \to a}(\{ \eta_{b \to i} \}_{b \in 
\partial i \setminus a}   ) \ , \qquad 
f_{i \to a}(\{ \eta_{b \to i} \})  (\sigma_i) =
\frac{1}{z_{i \to a}(\{ \eta_{b \to i} \})}
\prod_{b \in \partial i \setminus a} \eta_{b \to i}(\sigma_i) \ ,
\label{eq_bp_itoa}
\end{eqnarray}
where the functions $z$ are fixed by the normalization of the
$\eta$'s.  As we consider a tree factor graph these equations have a
unique solution, easily determined in a single sweep of updates from
the leaves of the graph towards its inside. Moreover the free entropy
of the model follows from this solution and reads
\begin{equation}
N \phi = \log Z = - \sum_{(i,a)} \log z_{ia}(\eta_{a \to i},\eta_{i \to a}) 
+  \sum_{a} \log z_a(\{\eta_{i \to a}\}_{i \in \partial a}  )
+  \sum_{i} \log z_i(\{\eta_{a \to i}\}_{a \in \partial i} ) \ .
\label{eq_Bethe}
\end{equation}
Here the first sum runs over the undirected edges of the factor graph
and the $z$'s are given by
\begin{equation}
z_{ia}= \sum_{\sigma_i} \eta_{a \to i}(\sigma_i) \eta_{i \to a}(\sigma_i) 
\ , \qquad
z_a = 
\sum_{\us_\da} w_a(\us_\da) \prod_{i \in \partial a} \eta_{i \to a}(\sigma_i) 
\ , \qquad
z_i = \sum_{\sigma_i} \prod_{a \in \partial i} \eta_{a \to i}(\sigma_i) \ .
\label{eq_def_zs}
\end{equation}
This computation is correct only on tree factor graphs.  Nevertheless
it is expected to yield good estimates of the marginals and free
entropy for a number of models on locally tree-like graphs.  The
belief propagation (BP) algorithm consists in iterating
Eqs.~(\ref{eq_bp_atoi},\ref{eq_bp_itoa}) in order to find an
(approximate) fixed point.  In particular, whenever the {\sf RS}
scenario holds, there should be one approximate solution of the above
equations that yields the correct leading order of the free entropy
density in the thermodynamic limit.  In any case, when dealing with
random factor graphs, one can always turn this simple computation into
a probabilistic one, defining a distribution of random messages by
reading (\ref{eq_bp_atoi},\ref{eq_bp_itoa}) in a distributional sense
with random weight functions and variables' degrees. The RS estimate
of the average free entropy is then obtained by averaging the various
terms in (\ref{eq_Bethe}) with respect to these random messages.

This approach can be refined in {\sf d1RSB} and {\sf 1RSB}
regimes. The BP equations (\ref{eq_bp_atoi},\ref{eq_bp_itoa}) should
be approximately valid if one computes the messages $\eta_{a \to i}$
and $\eta_{i \to a}$ as marginal laws of the measure $\mu_\gamma$
restricted to a single pure state $\gamma$.  When the number of pure
states is very large, one considers a distribution (with respect to
the pure states $\gamma$ with their weights $W_\gamma$) of messages on
each directed edge of the factor graph.

A simple and suggestive derivation of the 1RSB equations goes as follows.
Assume that the factor graph is a tree, and choose a subset $B$ of the
variable nodes that will act as a boundary, for instance (but not
necessarily) the leaves of the factor graph.  Each configuration
$\us_B$ of the variables in $B$ induces a conditional distribution
$\mu^{\us_B}$ on the remaining variables,
\begin{equation}
\mu^{\us_B}(\ut) = \frac{1}{Z^{\us_B}} w(\ut) \I(\ut_B = \us_B) \ ,
\end{equation}
where here and in the following $\I$ denotes the indicator function of
an event, and the normalizing factor $Z^{\us_B}$ is the partition
function restricted to the configurations coinciding with $\us_B$ on
the boundary.

Since the factor graph corresponding to $\mu^{\us_B}$ is still a tree,
the corresponding marginals and partition function $Z^{\us_B}$ can be
computed iterating the message passing equations
(\ref{eq_bp_atoi},\ref{eq_bp_itoa}), with an appropriate prescription
for the messages $\eta_{i \to a}$ emerging from variables $i \in B$,
namely $\eta_{i \to a}(\t_i) = \delta_{\s_i,\t_i}$. Let us denote by
$\eta_{a \to i}^{\us_B}$ and $\eta_{i \to a}^{\us_B}$ the
corresponding set of messages, solutions of
(\ref{eq_bp_atoi},\ref{eq_bp_itoa}) on all edges of the factor
graph. Further define, for $m\in\reals$, a probability measure on the
boundary conditions as
\begin{equation}
\tilde{\mu}(\us_B) = \frac{ (Z^{\us_B})^m }{\sum_{\us'_B} (Z^{\us'_B})^m} \ .
\label{eq_mu_tilde}
\end{equation}
The idea is to mimic the pure states of a large, loopy factor graph
model, by the boundary configurations of a tree model.  Calling $P_{a
  \to i}$ (resp.\ $P_{i \to a}$) the distribution of the messages
$\eta_{a \to i}^{\us_B}$ (resp.\ $\eta_{i \to a}^{\us_B}$) with respect
to $\tilde{\mu}$~\footnote{more precisely, with respect to the measure
  $\tilde{\mu}_{a\to i}$ (resp.\ $\tilde{\mu}_{i\to a}$) defined
  similarly for the factor graph $G_{a \to i}$ (resp.\ $G_{i \to
    a}$).}, a short reasoning reveals that
\begin{eqnarray}
P_{a \to i} (\eta) &=& \frac{1}{Z[\{ P_{j \to a} \},m ]}
\int\! \prod_{j \in \partial a \setminus i}\de P_{j \to a}(\eta_{j \to a})
\; \delta(\eta - f_{a\to i}(\{\eta_{j \to a} \})) 
\ z_{a\to i}(\{\eta_{j \to a} \})^m \ ,
\label{eq_spm_atoi} 
\\
P_{i \to a} (\eta) &=& \frac{1}{Z[\{ P_{b \to i} \},m ]}
\int\! \prod_{b \in \partial i \setminus a} \de P_{b \to i}(\eta_{b \to i})
\; \delta(\eta - f_{i\to a}(\{\eta_{b \to i} \})) 
\ z_{i \to a}(\{\eta_{b \to i} \})^m \ ,
\label{eq_spm_itoa}
\end{eqnarray}
where the functions $f$ and $z$ are defined in Eq.~(\ref{eq_bp_atoi}),
(\ref{eq_bp_itoa}), and the $Z[\cdots]$ are normalizing factors
determined by the condition $\int \de P_{a\to i}(\eta) = \int \de
P_{i\to a}(\eta)=1$.  Equations (\ref{eq_spm_atoi}),
(\ref{eq_spm_itoa}) coincide with the standard 1RSB equations with
Parisi parameter $m$~\cite{MePa}. In addition the free entropy density
associated to the law $\tilde{\mu}$, $N\Phi(m) \equiv \log\{
\sum_{\us_B} (Z^{\us_B})^m\}$ can be shown to be
\begin{equation}
N \Phi(m) = - \sum_{(i,a)\in E} \log Z_{ia}[P_{a \to i},P_{i \to a},m] 
+  \sum_{a\in F} \log Z_a[\{P_{i \to a}\}_{i \in \partial a},m]
+  \sum_{i\in V} \log Z_i[\{P_{a \to i}\}_{a \in \partial i},m] \ ,
\label{eq_Bethe_Phi}
\end{equation}
where the factors $Z_{\cdots}$ are fractional moments of the ones
$z_{\cdots}$ defined in Eq.~(\ref{eq_def_zs}), namely
\begin{equation}
Z_{ia} = \int \!\de P_{a \to i}(\eta_{a \to i}) \de P_{i \to a}(\eta_{i \to a})
\ z_{ia}^m  \ , \qquad
Z_a = \int \!\prod_{i \in \partial a}\de P_{i \to a}(\eta_{i \to a}) \ z_a^m
\ , \qquad
Z_i = \int \!\prod_{a \in \partial i}\de P_{a \to i}(\eta_{a \to i}) \ z_i^m
\ .
\end{equation}

As in the RS case, one can heuristically apply
(\ref{eq_spm_atoi},\ref{eq_spm_itoa}) on any graph, even if it is not
a tree.  Of particular interest is the limit $B\to\emptyset$.
Equations (\ref{eq_spm_atoi}), (\ref{eq_spm_itoa}) may have two
behaviors in this limit: $(i)$ All the distributions $P_{i\to a}$,
$P_{a\to i}$ become Dirac deltas in this limit.  In this case a `far
away' boundary has small influence on the system, and it is easily
seen by comparing (\ref{eq_Bethe}) and (\ref{eq_Bethe_Phi}) that
$\Phi(m) = m \phi$.  $(ii)$ These distributions remain non-trivial in
the limit $B\to \emptyset$. This case is interpreted as a consequence
of the existence of many pure states. In this situation, even a small
boundary influences the system by selecting one of such states. We
thus interpret the $B=\emptyset$ limit of $\Phi(m)$ as an estimate of
the replicated potential (\ref{eq_def_Phi}).

In Sec.~\ref{sec_reminder_rsb} we emphasized the special role played
by the value $m=1$: the dynamical transition is signaled by the
appearance of a non-trivial solution of the 1RSB equations with
$m=1$. This is particularly clear in the present derivation of the
1RSB equations. Indeed, the distribution $\tilde{\mu}$ of the boundary
condition coincides in this case with the Boltzmann distribution
$\mu$.

The existence of a non-trivial solution of the 1RSB equations at $m=1$
is thus related to a peculiar form of long range correlations under
$\mu$, as first pointed out in~\cite{MeMo}.  Such correlations can be
measured through a point-to-set correlation
function~\cite{BiBo,MaSiWe,CaGiVe}. For concreteness let us give an
expression of this correlation in the case of Ising spins.  Given a
variable node $i$ and a set of variable nodes $B$, we let
\begin{equation}
C(i,B) \equiv
\sum_{\us_B} \mu(\us_B) \left(\sum_{\s_i}\mu(\s_i|\us_B)\, \s_i \right)^2
- \left(\sum_{\s_i} \mu(\s_i)\, \s_i \right)^2 \ .\label{eq:Correlation}
\end{equation}
The reader will recognize the analogy between this expression and the
difference $q_1 - q_0$ of intra and inter-state overlaps~\cite{FrPa}.
The Boltzmann measure has long range point-to-set correlations if
$C(i,B)$ does not decay to $0$ when $d(i,B)$ grows.  Such correlations
were shown in~\cite{MosselHyperbolic,MoSe2} to imply a diverging
relaxation time.
%
%
\subsection{Application to constraint satisfaction problems}
\label{sec_reminder_csp}

This short overview of the cavity method did not rely on any
hypothesis on the form of the weight factors $w_a$ in
Eq.~(\ref{eq:Factor}).  We now comment briefly on the way this general
formalism is applied to constraint satisfaction problems (CSP), in
order to clarify the relationship of the present work with previous
studies. In a CSP the factors $a$ correspond to constraints, which can
be either satisfied or not by the configuration of their adjacent
variables, $\us_{\da}$. For a satisfiable instance of a CSP one can
take $w_a$ to be the indicator function of the event `constraint $a$
is satisfied.' Then the law defined in (\ref{eq:Factor}) is the
uniform distribution over the solutions of the CSP, the partition
function counts the number of such solutions and the free entropy
reduces to the logarithm of the number of solutions. This ``entropic''
method~\cite{MePaRi} is the most adequate to the study of the
satisfiable phase.

This approach is however ill-defined for unsatisfiable instances.  The
usual way to handle this case is to define a cost function $E(\us)$ on
the space of configurations, equal to the number of unsatisfied
constraints under the assignment $\us$. Following the traditional
notations of statistical mechanics one introduces an inverse
temperature $\beta$ and weighs the configurations with
$w(\us)=\exp[-\beta E(\us)]$. Small temperatures (large $\beta$)
favor low-energy configurations, in the limit $\beta \to \infty$ the
measure $\mu$ concentrates on the optimal configurations which
maximizes the number of satisfied constraints.  Let us detail
this approach which was originally followed
in~\cite{MePa_T0,MeZe,MeMeZe}. At the 1RSB level the pure states are
characterized by their energy density $e$ and their entropy density
$s$, with the free entropy density given by $\phi=s-\beta e$.
Defining the complexity $\Sigma(s,e)$ according to the number of pure
states with these two characteristics, Eq.~(\ref{eq_Legendre}) becomes
\begin{equation}
\Phi(\beta,m) = \sup_{s,e} [ \Sigma(s,e) + m (s-\beta e)] \ .
\end{equation}
If one takes now the limit $\beta \to \infty$ and assume $e>0$, the
entropic term becomes irrelevant; to obtain a finite result one has to
take at the same time $m\to 0$ such that the product $\beta m$,
usually denoted $y$, remains finite. One thus obtains
\begin{equation}
\Phi_{\rm e}(y) = \sup_e [\Sigma_{\rm e}(e)- y e] \ , \;\;\;\;\;
\Sigma_{\rm e}(e) \equiv
\sup_s \Sigma(s,e)\, .
\end{equation} 
In the unsatisfiable phase, the `energetic' cavity approach allows to
characterize the minimal energy of the problem.

In the case of satisfiable problems, one has to perform a second limit
$y \to \infty$ (after $\beta \to \infty$) to concentrate on the pure
states with $e=0$. It follows that the complexity thus computed is
$\sup_s \Sigma(s,e=0)$, i.e.\ the maximum of the entropic
complexity. In other words the procedure $y \to \infty$ after $\beta
\to \infty$ is equivalent to perform the entropic computation with a Parisi
parameter $m=0$, i.e.\ to weigh all the pure states in a same way,
irrespectively of their sizes. This is not a problem for the
determination of the satisfiability threshold $\alpha_{\rm s}$, which
corresponds to the disappearing of all zero-energy pure states, hence
to the vanishing of the maximal complexity $\Sigma(m=0)$. However the
value of $\alpha_{\rm d}$ in~\cite{MeZe,MeMeZe} corresponds to the
appearance of a solution of the 1RSB equations with $m=0$, and not
with $m=1$ which we argued to be the relevant value for the definition
of $\alpha_{\rm d}$.

In the rest of the paper we shall follow the entropic cavity method,
i.e.\ we take (\ref{eq:Factor}) to be the uniform measure over the
solutions of the CSP under study and keep a finite value for the
Parisi parameter $m$.  Before entering the details of this approach on
the example of random $k$-satisfiability, let us mention 
that the existence of exponentially numerous 
pure states (called clusters in this context)
for some values of $\alpha$ and $k$, has been proved in
\cite{MoMeZe,AcRi}. An intrinsic limitation of these works was 
that clusters were defined by much stricter conditions than the one exposed
above (which thus implied limitations on $\alpha$, $k$). 
The consequences of the existence of a distribution of
cluster's sizes have also been investigated in a toy model
in~\cite{rcm}.

We should also emphasize that for the simpler CSP known as
XORSAT~\cite{xor_1,xor_2}, a precise characterization of the clusters
has been achieved through rigorous methods. A good part of the
phenomena studied in the present paper is however absent of this
simpler model. In particular all clusters of XORSAT have the same size
because of the linear structure of the constraints.

\section{The cavity method applied to the random $k$-sat problem}
\label{sec_cavity}

\subsection{Some definitions}
\label{sec_cavity_def}
In the application of the formalism to $k$-satisfiability, we use
$\sigma_i \in \cX = \{-1,+1\}$ to encode the Boolean variables. A
constraint $a$ on $k$ variables $\us_\da$ is satisfied by all the
$2^k$ configurations except one, let us call it $\uJ^a = \{J^a_i: \,
i\in\da\}$, in which all the literals of the clause are false. The
weight factors are thus defined as $w_a(\us_\da) = \I(\us_\da \neq
\uJ^a)$, the indicator function of the event ``clause $a$ is
satisfied.''

\begin{figure}
\includegraphics[width=7cm]{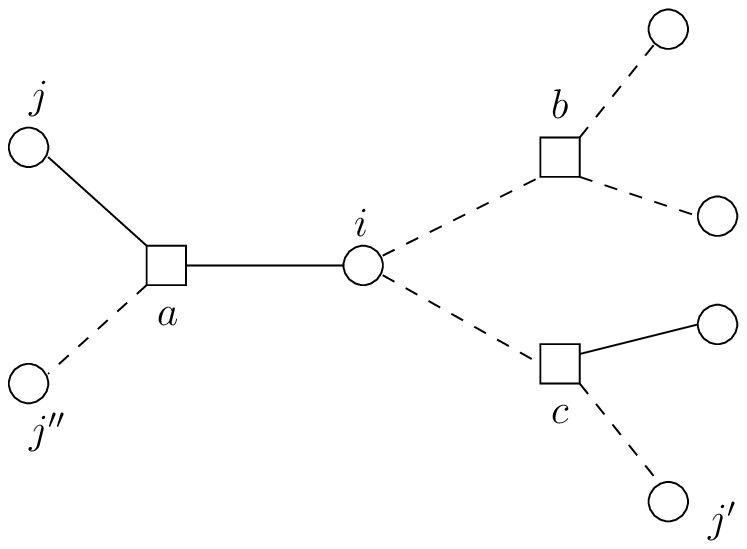}
\caption{An example of the factor graph representation of a
  satisfiability formula for $k=3$.  The values $J^a_i$ are encoded by
  drawing a solid (resp.\ dashed) edge between clause $a$ and variable
  $i$ if $\sigma_i=+1$ (resp.\ $-1$) satisfies clause $a$.  The
  distances between some of the variable nodes are
  $d_{i,j}=d_{i,j'}=d_{i,j''}=1$ and $d_{j,j'}=2$. The neighborhoods
  are for instance $\partial i =\{a,b,c\}$, $\partial a = \{i
  ,j,j''\}$, $\partial_+i=\{a\}$, $\partial_-i=\{b,c\}$,
  $\partial_+i(a)=\emptyset$, $\partial_-i(a)=\{b,c\}$,
  $\partial_+i(b)=\{c\}$, $\partial_-i(b)=\{a\}$.}
\label{fig_factor_graph}
\end{figure}

A formula is represented as a factor graph
(cf. Fig.~\ref{fig_factor_graph}) whose edges are labeled by $J^a_i$.
This suggests to refine the definition of the neighborhoods. Given a
variable node $i$, $\partial_+ i$ (resp.\ $\partial_- i$) will denote
the set of clauses which are satisfied by $\sigma_i=+1$
(resp.\ $\sigma_i=-1$). Further, given a clause $a\in\di$ we call
$\partial_+ i(a)$ (resp.\ $\partial_-i(a)$) the set of clauses in
$\partial i \setminus a$ which are satisfied by the same
(resp.\ opposite) value of $\sigma_i$ as is $a$.

For $k$-SAT formulas the general RS cavity
equations~(\ref{eq_bp_atoi}), (\ref{eq_bp_itoa}) can be written in a
pretty explicit form.  As the variables take only two values the
cavity probability messages $\eta_{a \to i}$ and $\eta_{i \to a}$ can
be parametrized by a single real number, that we shall call
respectively $u_{a \to i}$ and $h_{i \to a}$ and define by
\begin{equation}
\eta_{a\to i}(\sigma_i) = \frac{1- J_i^a \sigma_i \tanh u_{a \to i}}{2} 
\ , \qquad\qquad
\eta_{i\to a}(\sigma_i) = \frac{1- J_i^a \sigma_i \tanh h_{i \to a}}{2}  \ .
\label{eq_def_fields}
\end{equation}
With these conventions Eqs.~(\ref{eq_bp_atoi}), (\ref{eq_bp_itoa})
take the form
\begin{eqnarray}
u_{a \to i} &=& f(\{ h_{j \to a} \}_{j\in \partial a \setminus i} ) \ , \qquad
f(h_1,\dots,h_{k-1}) =
-\frac{1}{2} 
\log \left( 1-  \prod_{i=1}^{k-1} \frac{1-\tanh h_i}{2} \right)
\ , \label{eq_recurs_def1} \\
h_{i \to a} &=& \sum_{b \in \partial_+ i(a) } u_{b \to i}
- \sum_{b \in \partial_- i(a) } u_{b \to i} \ .
\label{eq_recurs_h}
\end{eqnarray}

We are interested in the regime where the number $M$ of uniformly
chosen clauses and the number of variables $N$ both diverge at fixed
ratio $\alpha = M/N$. The random factor graphs thus generated enjoy
properties reminiscent of the Erd\"os-R\'enyi random graphs
$G(N,M)$~\cite{rgraphs,MeMoBook}.  In particular, for a uniformly
random variable node $i$, the number of clauses in $\partial_+ i$ and
$\partial_- i$ converges to two i.i.d Poisson random variables of mean
$\alpha k/2$.  The same statement is true for $\partial_+i(a)$ and
$\partial_-i(a)$ when $(i,a)$ is an uniformly chosen edge of the
factor graph.  The degree distribution is a very local description of
a graph, looking at one node or edge only.  It is however easy to show
that any bounded neighborhood of a uniformly random node $i$ converges
to a random (Galton-Watson) tree with the same degree distribution
\cite{MeMoBook}.
%
%
\subsection{The RS description of the random formulae ensemble}

The replica-symmetric treatment of the random $k$-SAT problem was
first worked out using the replica formalism in~\cite{MoZe}. In the
cavity formulation one interprets the BP equations
(\ref{eq_bp_atoi},\ref{eq_bp_itoa},\ref{eq_recurs_def1},\ref{eq_recurs_h})
in a probabilistic way. More precisely, we introduce the distributions
of $u_{a\to i}$, $h_{i\to a}$ (over the choice of the random formula)
and denote them as $\P_{(0)}(h)$ and $\Q_{(0)}(u)$. These
distributions satisfy the distributional equations:
\begin{equation}
u \eqd f(h_1,\dots,h_{k-1}) \ , \qquad
h \eqd \sum_{i=1}^{l_+} u_i^+ - \sum_{i=1}^{l_-} u_i^- \ .
\label{eq_recurs_RS}
\end{equation}
In these expressions $h,\{h_i\}$ (resp.\ $u,\{u_i^\pm \}$) are
independent copies of the random variable of distribution
$\P_{(0)}(h)$ (resp.\ $\Q_{(0)}(u)$), the function $f$ is defined in
Eq.~(\ref{eq_recurs_def1}) and $l_\pm$ are two independent Poisson
random variables of mean $\alpha k/2$.  The symbol $\ed$ denotes
identity in distribution\footnote{More explicitly, given two random
  variables $X$ and $Y$ we write $X\ed Y$ if the distributions of $X$
  and $Y$ coincide. For instance, if $X, X_1, X_2$ are iid standard
  normal random variables, $X \ed (X_1+X_2)/\sqrt{2}$}.

The RS prediction for the entropy reads
\begin{equation}
\phi_{(0)} = - \alpha k \E\, \log z_1(u,h) + \alpha \E\, 
\log z_2(h_1,\dots,h_k)
+ \E\, \log z_3(
u_1^+,\dots,u_{l_+}^+,u_1^-,\dots,u_{l_-}^-) \ ,
\label{eq_phi_RS}
\end{equation}
where the expectations are over i.i.d. copies of the random variables
$u$ and $h$, and $l_\pm$ are as above. The various entropy shifts are
obtained by rewriting the $z$'s in Eq.~(\ref{eq_def_zs}) in terms of
$u$ and $h$,
\begin{eqnarray}
&&z_1(u,h) = 1+\tanh h \tanh u \ , \\
&&z_2(h_1,\dots,h_k) = 1 - \prod_{i=1}^k \frac{1-\tanh h_i}{2}
\ , \\
&&z_3(
u_1^+,\dots,u_{l_+}^+,u_1^-,\dots,u_{l_-}^-
) =
\prod_{i=1}^{l_+} (1+\tanh u_i^+) 
\prod_{i=1}^{l_-} (1-\tanh u_i^-) +
\prod_{i=1}^{l_+} (1-\tanh u_i^+)
\prod_{i=1}^{l_-} (1+\tanh u_i^-) \ .
\label{eq_def_z3}
\end{eqnarray}
Similarly the RS overlap can be computed as
\begin{equation}
q_0 = \E[\tanh^2 h] \ .
\end{equation}

Several equivalent expressions of the RS entropy can be found in
the literature; the choice we made in (\ref{eq_phi_RS}) has the
advantage of being variational. By this we mean that the stationarity
conditions of the function $\phi_{(0)}[\P,\Q,\alpha]$ with respect to
$\P$ and $\Q$ are nothing but the self-consistency equations
(\ref{eq_recurs_RS}).  Note also that the rigorous results
of~\cite{FrLe,PaTa} imply that\footnote{In \cite{FrLe,PaTa} this claim is
  made for $k$ even. However the proof holds verbatim for $k$ odd as
  well. To the best of our knowledge, this was observed first by
  Elitza Maneva in 2005.} the entropy density $\phi$ is upper-bounded
by the RS $\phi_{(0)}$ for any trial distribution $\P$, as long as
$\Q$ is linked to $\P$ by the first equation in (\ref{eq_recurs_RS}),
for a regularized version of the model at finite temperature.
Moreover the RS description was proven to be valid for small values of
$\alpha$ in~\cite{MoSh}.

The numerical resolution of the equation on the order parameter is
relatively easy. The distributions $\P_{(0)}$ and $\Q_{(0)}$ can
indeed be represented by samples (or populations) of a large number
$\cal N$ of representatives, $\{h_i\}_{i=1}^{\cal N}$ and
$\{u_i\}_{i=1}^{\cal N}$. The fixed point condition stated in
(\ref{eq_recurs_RS}) is looked for by an iterative population dynamics
algorithm~\cite{popu,MePa,MeMoBook}.

We turn now to the cavity formalism at the 1RSB level, which assumes
the organization of pure states described in Section
\ref{sec_reminder_rsb}.
%
%
\subsection{The 1RSB description of the random formulae ensemble}

As in the RS case, when the underlying formula is random, the messages
$P_{i\to a}$, $P_{a\to i}$ along a uniformly random edge become random
variables, whose distributions are denoted as $\P_{(1)}[P]$,
$\Q_{(1)}[Q]$.  These distributions satisfy a couple of distributional
equations, that are the probabilistic version of
Eqs.~(\ref{eq_spm_atoi},\ref{eq_spm_itoa}),
\begin{eqnarray}
Q(\bullet) & \eqd & 
\frac{1}{\Z_4[P_1,\dots,P_{k-1}]}
\int \prod_{i=1}^{k-1} \de P_i(h_i) \ \delta\left(
\bullet -f(h_1,\dots,h_{k-1}) \right)  
z_4(h_1,\dots,h_{k-1})^m \ ,\label{eq_1RSB_Q} \\
P(\bullet) &\eqd& \frac{1}{\Z_3[
\{Q_i^+\},\{Q_i^-\}
]}
\int \prod_{i=1}^{l_+} \de Q_i^+(u_i^+) \prod_{i=1}^{l_-} \de Q_i^-(u_i^-) \ 
\delta\left(\bullet - \sum_{i=1}^{l_+} u_i^+ + \sum_{i=1}^{l_-} u_i^- \right)
z_3(
\{u_i^+\}_{i=1}^{l_+},\{u_i^-\}_{i=1}^{l_-}
)^m \ ,
\label{eq_1RSB_P}
\end{eqnarray}
where the $P$'s (resp.\ $Q$'s) are i.i.d. from $\P_{(1)}$
(resp.\ $\Q_{(1)}$) and $l_\pm$ have the above stated Poissonian
distribution. The entropy shift $z_3$ used in Eq.~(\ref{eq_1RSB_P})
was defined in Eq.~(\ref{eq_def_z3}), while $z_4$ is given by
\begin{equation}
z_4(h_1,\dots,h_{k-1}) = 2 - \prod_{i=1}^{k-1} 
\frac{1-\tanh h_i}{2} = 1+e^{-2f(h_1,\dots,h_{k-1})} \ .
\label{eq_def_z4}
\end{equation}

Finally, the 1RSB potential is obtained by taking the expectation of
Eq.~(\ref{eq_Bethe_Phi}). One gets
\begin{equation}
\Phi(m) = - \alpha k \E\,\log \Z_1[Q,P] + \alpha \E\, 
\log \Z_2[P_1,\dots,P_k]
+ \E\, \log \Z_3[Q_1^+,\dots,Q_{l_+}^+,Q_1^-,\dots,Q_{l_-}^-] \ ,
\label{eq_1RSB_Phi}
\end{equation}
where the factors $\Z_i$ are weighted averages of the corresponding
entropy shifts,
\begin{eqnarray}
&&\Z_1[Q,P] = \int \de P(h)\de Q(u) \ z_1(u,h)^m  \ , \\
&&\Z_2[P_1,\dots,P_k] = \int \prod_{i=1}^k \de P_i(h_i) \ z_2(h_1,\dots,h_k)^m
\ , \\
&&\Z_3[Q_1^+,\dots,Q_{l_+}^+,Q_1^-,\dots,Q_{l_-}^-] =
\int \prod_{i=1}^{l_+} \de Q_i^+(u_i^+) \prod_{i=1}^{l_-} \de Q_i^-(u_i^-)
\ z_3(u_1^+,\dots,u_{l_+}^+,u_1^-,\dots,u_{l_-}^-)^m
\ .
\end{eqnarray}
The inter and intra-state overlaps are given, respectively, by
\begin{equation}
q_0 = \E\left[ \left( \int \de P(h) \tanh h \right)^2  \right] \ , \qquad
q_1 = \E\left[ \int \de P(h) \tanh^2 h  \right] \ .
\label{eq_def_overlaps}
\end{equation}

The variational property discussed at the RS level still applies to
the 1RSB potential. This is of particular interest for the computation
of the internal entropy of the states, given by a derivative with
respect to $m$. This derivation can be applied to the explicit
dependence only, and yields
\begin{eqnarray}
\phi_{\rm int}(m) =
&-&\alpha k \E\left[\frac{\int\!\de P(h) \de Q(u) \ 
z_1(u,h)^m \log z_1(u,h)}
{\Z_1(Q,P)}\right] \nonumber \\
&+& \alpha \E\left[\frac{\int \prod_{i=1}^k \!\de P_i(h_i) \ 
z_2(\{h_i\}_{i=1}^k)^m 
\log z_2(\{h_i\}_{i=1}^k) }{\Z_2[\{P_i\}_{i=1}^k] } \right] \nonumber \\
&+&\E\left[ \frac{ 
\int \prod_{i=1}^{l_+} \!\de Q_i^+(u_i^+) \prod_{i=1}^{l_-} \!\de Q_i^-(u_i^-)
\ z_3(\{u_i^+\}_{i=1}^{l_+},\{u_i^-\}_{i=1}^{l_-})^m  
\log z_3(\{u_i^+\}_{i=1}^{l_+},\{u_i^-\}_{i=1}^{l_-}) }
{\Z_3[\{Q_i^+\}_{i=1}^{l_+},\{Q_i^-\}_{i=1}^{l_-} ]} \right] \ .
\label{eq_phi_int}
\end{eqnarray}
The rigorous results of~\cite{FrLe,PaTa} also imply $\phi \le
\Phi(m)/m$ for any value of $m$ in $(0,1)$, and any trial order
parameter $\P$ (with $\Q$ defined by Eq.~(\ref{eq_1RSB_Q})).

The numerical resolution of the 1RSB equations
(\ref{eq_1RSB_Q},\ref{eq_1RSB_P}) is in general much harder than the
one of their RS counterparts (compare with
Eq.~(\ref{eq_recurs_RS})). The population dynamics algorithm
represents $\P_{(1)}$ by a sample of distributions $\{ P_i
\}_{i=1}^{\cal N}$, which themselves have to be encoded, for each $i$,
by a finite set of cavity fields $\{h_{i,j}\}_{j=1}^{{\cal N}'}$.
This drastically limits the sizes $\cal N$ and ${\cal N}'$, and hence
the precision of the numerical results. Moreover generating one
element, say $Q_i$, from $k-1$ $P_i$'s is by itself a non trivial
task. The various fields representing $Q_i$ are weighted in a non
uniform way because of the factor $z_4^m$ in Eq.~(\ref{eq_1RSB_Q}),
which forces the use of delicate resampling procedures.

These equations can be greatly simplified analytically for two particular
values of $m$, namely 0 and 1. For the sake of readability we postpone
the discussion of these important simplifications until
Section~\ref{sec_1rsb_tricks}, and proceed in the next section with
the presentation and the interpretation of the results obtained either
at arbitrary $m$ with the full numerical procedure (whose
implementation details are exposed in Appendix~\ref{sec_numerics}) or
in $m=0,1$ with the simplified, more precise ones.

\section{Transitions in the satisfiable regime of random $k$-sat}
\label{sec_transitions}

\subsection{The dynamical, condensation and satisfiability transitions 
for $k \ge 4$}
\label{sec:res_k4}

\begin{figure}
\includegraphics[width=0.45\textwidth]{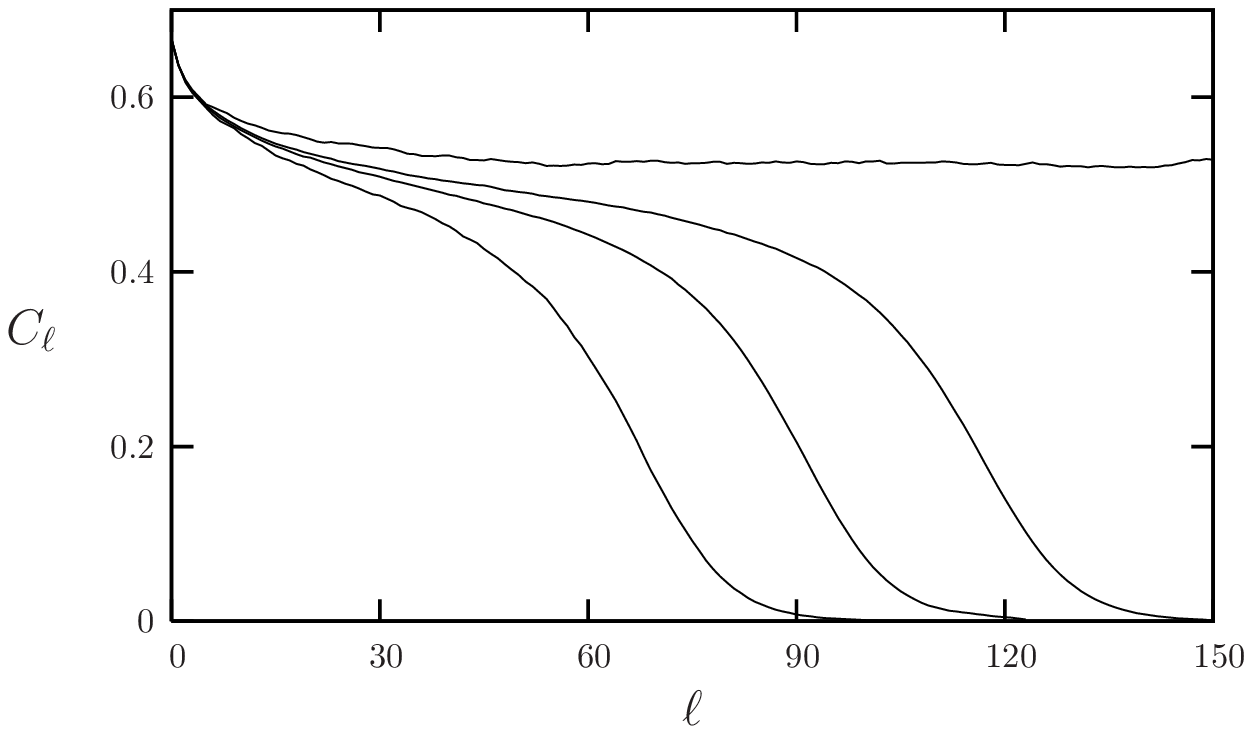}
\caption{The point-to-set correlation function for $k=4$, from left to
  right $\alpha=9.30$, $\alpha=9.33$, $\alpha=9.35$ and $\alpha=9.40$}
\label{fig:k4_C_ell}
\end{figure}

\begin{figure}
\includegraphics[width=0.6\textwidth]{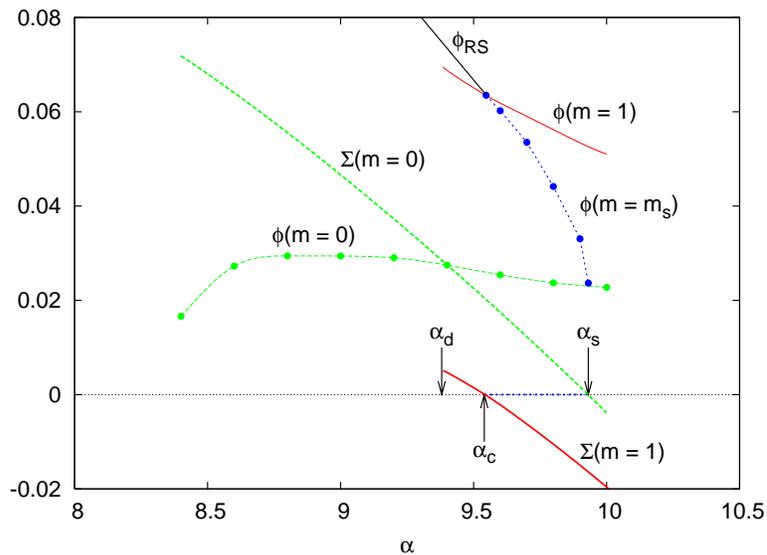}
\caption{The complexity $\Sigma$ and the internal entropy
  $\phi_{\rm int}$ for the values $m=0,1$, and $m=m_{\rm s}$ in the 
{\sf 1RSB} regime, for $k=4$.} 
\label{fig:k4-Sigma}
\end{figure}

Let us begin our discussion of the satisfiable regime of random
$k$-SAT by studying the case $k=4$, the values $k \ge 4$ having the
same qualitative behavior. On the other hand, the phenomenology of
$3$-SAT is different and we report on it in Sec.~\ref{sec:res_k3}.

Following the program of Sec.~\ref{sec_reminder} we first have to
determine the value $\alpha_{\rm d}$ for the appearance of a
non-trivial solution of the 1RSB equations with $m=1$. To this aim we
compute the point-to-set correlation function $C_\ell$, that is the
average of the correlation function (\ref{eq:Correlation}) between a
randomly chosen variable $i$ and the set $B$ of variables at distance
$\ell$ from it. The plots of Fig.~\ref{fig:k4_C_ell} show that for
$\alpha \le \alpha_{\rm d} \approx 9.38$ this correlation vanishes at
large distance, while for larger values of $\alpha$ a strictly
positive long range correlation sets in discontinuously.  To
distinguish between the {\sf d1RSB} and {\sf 1RSB} regime we then
compute the complexity $\Sigma(m=1)$. As demonstrated in
Fig.~\ref{fig:k4-Sigma} this is strictly positive at $\alpha_{\rm d}$,
then decreases continuously until it vanishes at $\alpha_{\rm
  c}\approx 9.547$.  Finally the satisfiability transition
$\alpha_{\rm s}$ is found from the criterion of vanishing of
$\Sigma(m=0)$, i.e.\ the maximum of the entropic complexity curve (see
Fig.~\ref{fig:k4-Sigma}): the value $\alpha_{\rm s}\approx 9.931$ is
in agreement with~\cite{MeMeZe} and we shall show in
Sec.~\ref{sec_1rsb_m0} that this is indeed the same calculation.

To summarize, we find the three regimes {\sf RS}, {\sf d1RSB}, {\sf
  1RSB} described in Sec.~\ref{sec_reminder_rsb} occurring in this
order, for the values of $\alpha$ in $[0,\alpha_{\rm d}]$,
$[\alpha_{\rm d},\alpha_{\rm c}]$ and $[\alpha_{\rm c},\alpha_{\rm
  s}]$. We expect this pattern of transitions to be the same for all
$k\ge 4$. This is supported by our numerical investigations for
$k=4,5,6$ (see Tab.~\ref{tab_thresholds} for a summary of the
numerical values of the thresholds), and by the large-$k$ expansions
presented in Sec.~\ref{sec_large_k}.

The entropy density (see Fig.~\ref{fig:k4-Sigma}) is given by the RS
formula both in the {\sf RS} and {\sf d1RSB} regimes. In the latter
case it has to be understood as the sum of the complexity
$\Sigma(m=1)$ and of the internal entropy of the associated states,
$\phi_{\rm int}(m=1)$.  On the contrary for $\alpha \in [\alpha_{\rm
  c},\alpha_{\rm s}]$ it is necessary to compute the whole function
$\Sigma(\phi)$ by varying $m$.  The entropy density coincides with the
one of dominant clusters, and is given by the point where
$\Sigma(\phi)$ vanishes.

\begin{table}
\begin{center}
\begin{tabular}{| c | c | c | c | c |}
\hline
$k$ & $\alpha_{\rm d}$ & $\alpha_{\rm c}$ & 
$\alpha_{\rm s}$\cite{MeMeZe} &  $\alpha_{\rm f}$ \\
\hline
\hline
3 & 3.86 & 3.86 & 4.267  &   *   \\
\hline
4 &  9.38 & 9.547 & 9.931  &  9.88 \\
\hline
5 & 19.16 & 20.80 & 21.117 &   *   \\
\hline
6 & 36.53 & 43.08 & 43.37  & 39.87  \cite{rearr_csp}  \\
\hline
\hline
\end{tabular}
\end{center}
\caption{Numerical values of the various critical thresholds. 
  For $k=3$ we have formally $\alpha_{\rm c}=\alpha_{\rm d}$,
  see the text for details on the nature of the difference between
  $k=3$ and $k \ge 4$.}
\label{tab_thresholds}
\end{table}

\subsection{The entropic complexity curves}

\begin{figure}
\includegraphics[width=0.6\textwidth]{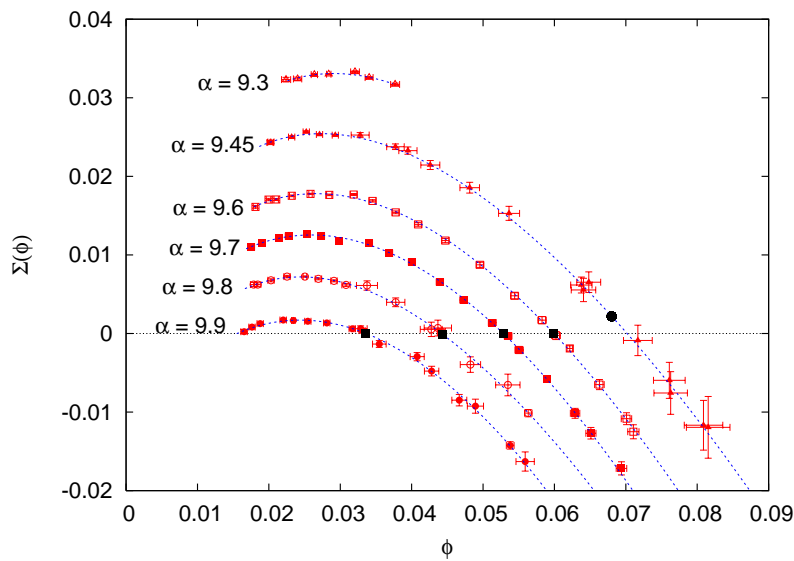}
\caption{The complexity $\Sigma(\phi)$ for $k=4$ and several values of 
$\alpha$: from top to bottom $\alpha=9.3$, $9.45$, $9.6$, $9.7$, $9.8$
and $9.9$.}
\label{fig:k4-Sigma_phi}
\end{figure}

The curves $\Sigma(\phi)$ are shown in Fig.~\ref{fig:k4-Sigma_phi} for
several values of $\alpha$. The symbols are obtained in a parametric way,
by solving the 1RSB equations for various values of $m$ and plotting the
point $(\phi_{\rm int}(m),\Sigma(m))$.
The lines in Fig.~\ref{fig:k4-Sigma_phi}
are numerical interpolations, obtained by fitting not directly
$\Sigma(\phi)$, but instead the data for $\Phi(m)$ with a
generic smooth function\footnote {We have tried different fitting
  functions and all provide equivalent and very good results thanks to
  the smoothness of $\Phi(m)$.} and then analytically deriving
the fitting function to obtain the curves in
Fig.~\ref{fig:k4-Sigma_phi}. The agreement of this fitting procedure 
with the parametric plot is excellent.  
The three regimes are clearly illustrated on this figure:
\begin{itemize}
\item For $\alpha < \alpha_{\rm d}$ a portion of the curve
  $\Sigma(\phi)$ can exist (for instance there is a solution of the
  1RSB equation with $m=0$ for $\alpha \ge 8.297$~\cite{MeMeZe}), yet
  it has no point of slope $-m=-1$.  The contribution of these
  clusters is negligible compared to the dominant RS cluster.
\item For $\alpha \in [\alpha_{\rm d},\alpha_{\rm c}]$ (see e.g.\
  $\alpha=9.45$ data in Fig.~\ref{fig:k4-Sigma_phi}) the complexity
  $\Sigma(m=1)$ exists and is positive (it is marked by a black circle
  in the figure).
\item For $\alpha \in [\alpha_{\rm c},\alpha_{\rm s}]$ (see e.g.\
  $\alpha=9.6,9.7,9.8,9.9$ in Fig.~\ref{fig:k4-Sigma_phi}) the
  complexity $\Sigma(m=1)$ is negative and thus the $\Sigma(\phi)$
  curve vanishes at $\phi(m_{\rm s})$ (marked with a black square),
  where the slope (in absolute value) is smaller than 1 and equals
  $m_{\rm s}(\alpha)$. The measure is dominated by a subexponential
  number of clusters of entropy $\phi(m_{\rm s})$, shown as a function
  of $\alpha$ in Fig.~\ref{fig:k4-Sigma}.
\end{itemize}

\begin{figure}
\includegraphics[width=0.6\textwidth]{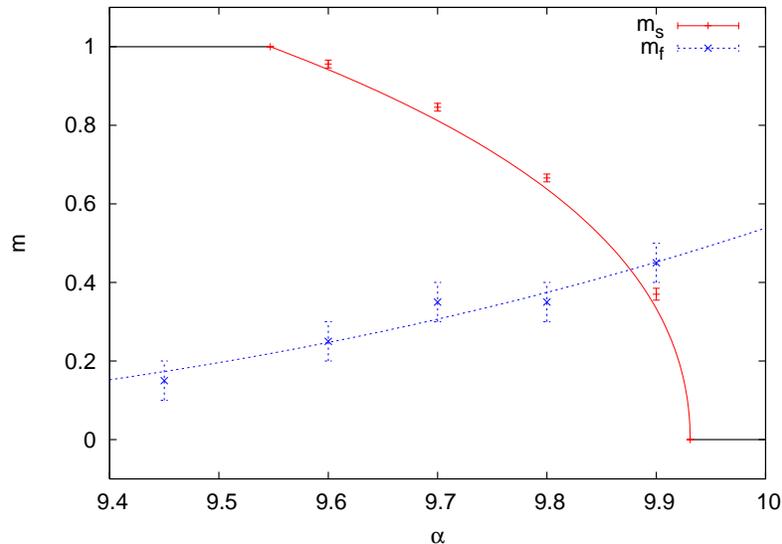}
\caption{The value of the Parisi parameter $m_{\rm s}$ in the
  thermodynamically relevant pure states of the {\sf 1RSB} regime in random
  4-SAT, and the freezing transition $m_{\rm f}$.}
\label{fig:k4-ms}
\end{figure}

The value thus estimated of the Parisi parameter $m_{\rm s}(\alpha)$
in the {\sf 1RSB} regime is plotted in Fig.~\ref{fig:k4-ms} (it is
identical to 1 in the {\sf d1RSB} region).  The curve close to the
$m_{\rm s}$ data is not a fit, but instead an explicit approximate
expression for $m_{\rm s}(\alpha)$ which becomes exact in the large $k$ 
limit (see Sec.~\ref{sec_large_k} for details).
Indeed Eq.~(\ref{eq_m_largek}) (valid to leading order at large $k$)
can be equivalently rewritten as
\begin{equation}
\frac{\alpha_{\rm s} - \alpha}{\alpha_{\rm s} - \alpha_{\rm c}} =
\frac{1-2^m(1-m \log 2)}{2\log 2 - 1}\;,
\label{eq:ms}
\end{equation}
and this gives an expression for $m_{\rm s}(\alpha)$ once values of 
$\alpha_{\rm c}$ and $\alpha_{\rm s}$ determined numerically 
for $k=4$ are plugged
into Eq.~(\ref{eq:ms}). Note that the solution to
Eq.~(\ref{eq:ms}) is such that $(i)$ $m_{\rm s}(\alpha_{\rm c}) = 1$,
$(ii)$ $m_{\rm s}(\alpha_{\rm s}) = 0$ and $(iii)$ $m_{\rm s}$
vanishes as a square root at $\alpha_{\rm s}$. The finite $k$ corrections
to the expression (\ref{eq:ms}) seem already small for $k=4$, as can be
inferred from the good agreement with the numerical data displayed in
Fig.~\ref{fig:k4-ms}. This fact was also noticed for the coloring problem 
in~\cite{FlLe}.

\begin{figure}
\includegraphics[width=0.6\textwidth]{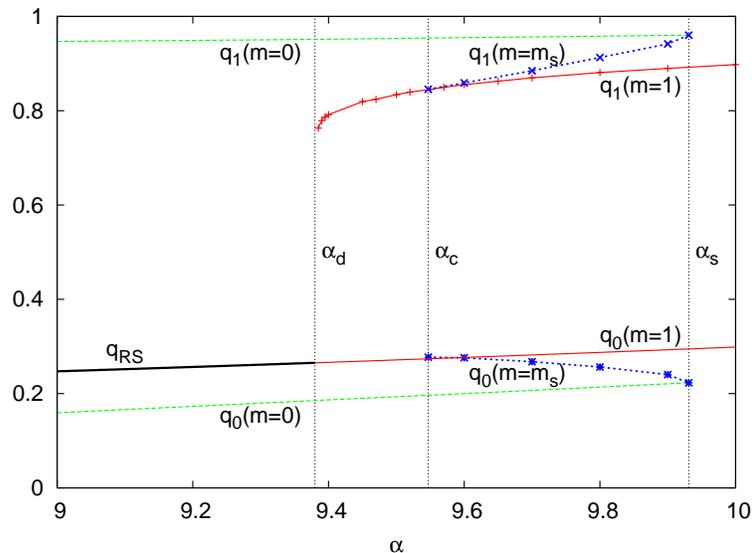}
\caption{Intra and inter-state overlaps for $k=4$.}
\label{fig:k4-q0q1}
\end{figure}

Once we compute the optimal value $m_{\rm s}$ for each value of
$\alpha$, we can plot in Fig.~\ref{fig:k4-q0q1} the overlap $q_0$ and
$q_1$ of the dominating clusters as a function of $\alpha$.  Notice
that the inter-state overlap $q_0$ is an increasing function of
$\alpha$ for any fixed value of $m$, but becomes a decreasing function
of $\alpha$ between $\alpha_{\rm c}$ and $\alpha_{\rm s}$ where we
take $m=m_{\rm s}(\alpha)$.

We did not attempt a complete determination of the portion of the
plane $(\alpha,m)$ where non-trivial solutions of the 1RSB equations
can be found. From our numerical investigations it seems that
solutions with smaller values of $m$ appear at smaller values of
$\alpha$, i.e.\ the threshold $\alpha_{\rm d}(m)$ is an increasing
function in the range of parameters we considered.  In particular,
solutions with negative $m$ appear at rather small values of $\alpha$. 
The limit of very large negative values of $m$ is
however difficult to study numerically, and more work could be done on
this issue; the corresponding pure states are tiny because their variables 
are overconstrained, which plagues the numerical resolution of the 1RSB
equations.

\subsection{On the presence of frozen variables in clusters of solutions}
\label{sec_frozenvar}

Another characterization of the clusters of solutions, besides their
internal entropy and self-overlap, is the presence or not of frozen
variables, that is variables that take the same value in all the
solutions of the cluster. In technical terms this corresponds to a
non-vanishing weight on $\pm \infty$ in the 1RSB cavity field
distributions $P(h)$ (see Eq.~(\ref{eq_def_hard}) below). Our data
show that, given a value of $\alpha$, there exists a threshold $m_{\rm
  f}(\alpha)$ such that clusters described by $m<m_{\rm f}$ do contain
frozen variables, while those with $m>m_{\rm f}$ do not. This is
consistent with the intuition: the freezing of variables is correlated
with a smaller value of the internal entropy, hence of $m$. Numerical
estimates for the line $m_{\rm f}(\alpha)$ are plotted in
Fig.~\ref{fig:k4-ms} for $k=4$. The large error bars are due to the
fact that we have checked the presence of frozen variables only at $m$ values 
which are multiples of $0.1$ (and no interpolation can be done in between,
since the property is just true or false).  The interpolating curve is a fit
to the $m_{\rm f}(\alpha)$ data with the function $A (x-8.297)^B$ (for
$m=0$ the critical value of $\alpha$ is $\alpha_{\rm d}(m=0) \approx
8.297$~\cite{MeMeZe}).  
The freezing transition $\alpha_{\rm f}$ is defined by the
appearance of frozen variables in dominating clusters, that is $m_{\rm
  s} (\alpha_{\rm f}) = m_{\rm f}(\alpha_{\rm f})$. From the crossing
of these two lines in Fig.~\ref{fig:k4-ms} we estimated the freezing
threshold for $k=4$ at $\alpha_{\rm f} \approx 9.88$.

The fact that the freezing transition occurs after the condensation
one for $k=4$ is not generic; for $k\ge 6$ the threshold $m_{\rm
  f}(\alpha)$ reaches 1 at $\alpha_{\rm f} \le \alpha_{\rm
  c}$~\cite{rearr_csp}, hence in a part of the {\sf d1RSB} regime the
dominating clusters do contain frozen variables for these values of
$k$.

Let us however emphasize that generally $\alpha_{\rm d} < \alpha_{\rm
  f}$, i.e.\ that in random $k$-satisfiability (and also in
$q$-coloring~\cite{FlLe}) clustering can occur without implying the
freezing of variables. This fact has been obscured up to now because
the energetic cavity method~\cite{MeZe,MePa_T0} focused precisely on
the fraction of frozen variables in the $m=0$ solution of the 1RSB
equations, and because in the simpler XORSAT model~\cite{xor_1,xor_2}
the freezing and clustering transitions coincide. We
refer the reader to~\cite{rearr_csp} for a more extensive study of the
freezing transition, in particular its interpretation in terms of the
divergence of the minimal rearrangements~\cite{MoSe} it induces, 
and to~\cite{AcRi} where it has been proven that
frozen variables exist in every cluster for $k \ge 9$ and $\alpha$
large enough.

\subsection{$k=3$, a special case}
\label{sec:res_k3}

\begin{figure}
\includegraphics[width=0.45\textwidth]{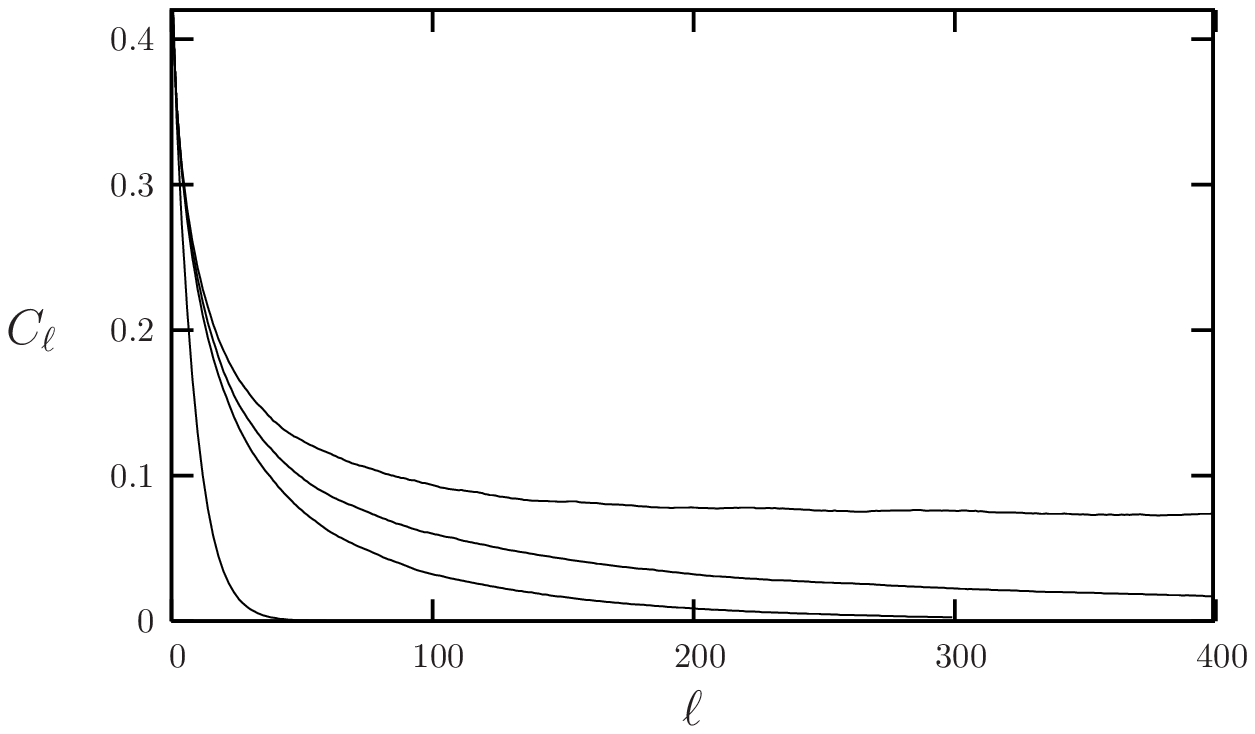}
\caption{The point-to-set correlation function for $k=3$, from left to
  right $\alpha=3.60$, $\alpha=3.84$, $\alpha=3.86$, $\alpha=3.88$.}
\label{fig:k3_C_ell}
\end{figure}

We turn now to the description of our numerical results in the
particular case $k=3$, recently investigated also in~\cite{k3_Zhou}.
The onset of long-range point to set correlations, displayed in
Fig.~\ref{fig:k3_C_ell} through the correlation function $C_\ell$, is
qualitatively different from $k=4$ (compare with
Fig.~\ref{fig:k4_C_ell}).  The long range correlation
$\lim_{\ell\to\infty}C_\ell$ grows indeed continuously from $0$ at
$\alpha_{\rm d}$ (in qualitative agreement with the variational
approximation of~\cite{BiMoWe}).  In fact this transition coincides
with a local instability of the RS solution with respect to 1RSB
perturbations (this is a generic fact for all models with continuous 
dynamic transitions).  
A numerical procedure can be used to locate precisely
this instability~\cite{PaPaRa,MoPaRi,CaKrRi}.  We get the estimate
$\alpha_{\rm d} = \alpha_{\rm stab} \approx 3.86$. Please note that
for $k \ge 4$ this local instability occurs after the discontinuous
transition, for instance at $\alpha_{\rm stab} \approx 10.2$ for $k=4$.

\begin{figure}
\includegraphics[width=0.6\textwidth]{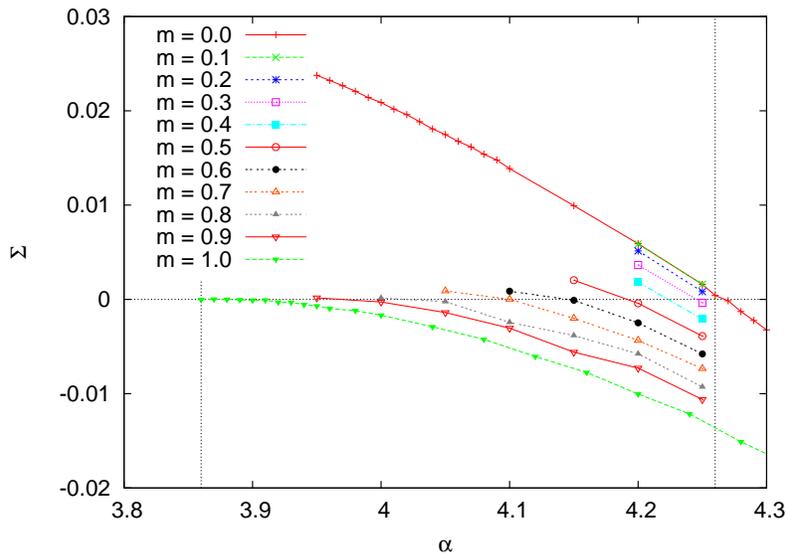}
\caption{The complexity $\Sigma$ for $k=3$ and $m$ from 0 (highest
  curve) to 1 (lowest curve).  For $0<m<1$ the domain of existence of
  $\Sigma$ may be slightly larger than the one shown in the plot (we
  have simulated only $\alpha$ values multiples of $0.05$).}
\label{fig:k3-Sigma}
\end{figure}

\begin{figure}
\includegraphics[width=0.6\textwidth]{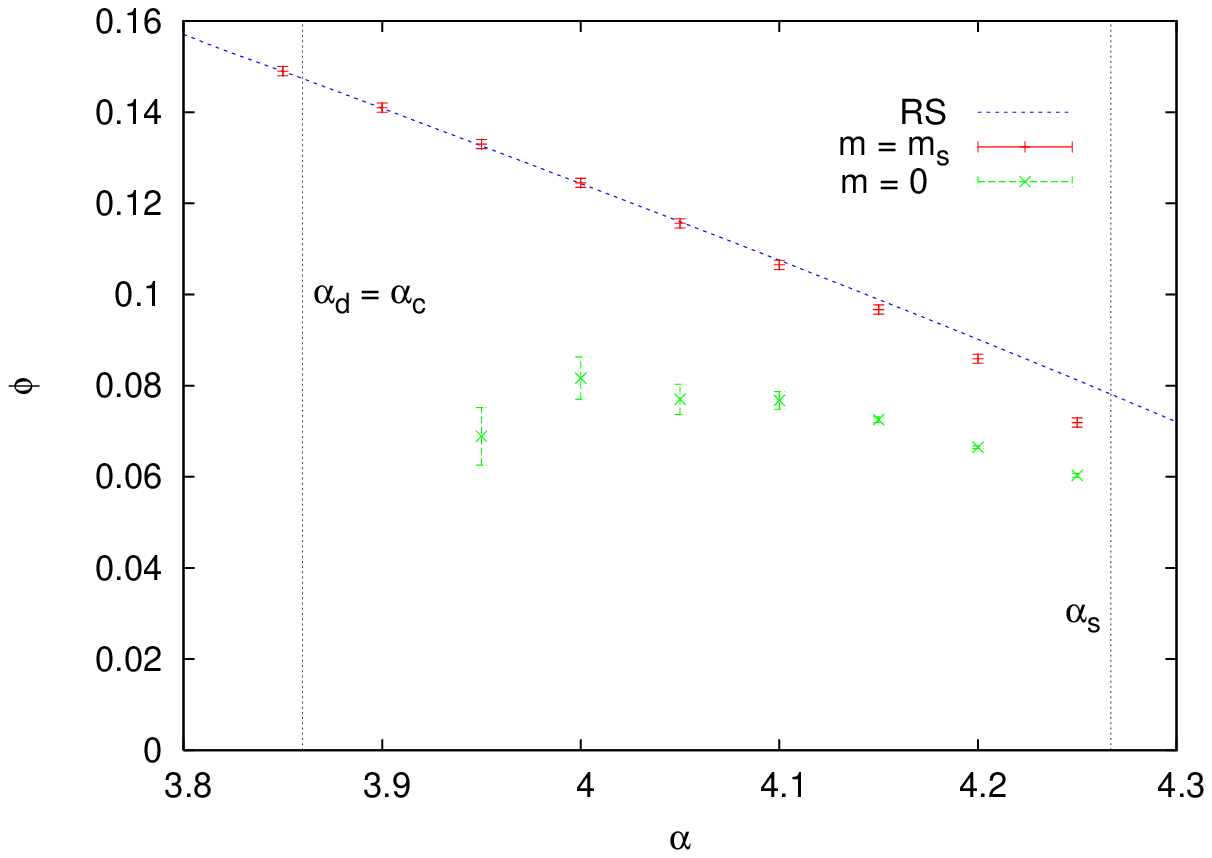}
\caption{The 1RSB estimate for the entropy of random 3-SAT, compared to the
  replica symmetric (RS) estimate and to the internal entropy of the $m=0$
  solution, corresponding to the maximum of the $\Sigma(\phi)$ curve.}
\label{fig:k3-phi_alpha}
\end{figure}

For $\alpha>\alpha_{\rm d}$ the complexity $\Sigma(m=1)$ decreases
continuously from 0 (see lowest curve in Fig.~\ref{fig:k3-Sigma}):
there is no {\sf d1RSB} regime for 3-SAT. We then turned to the
resolution of the 1RSB equations for other values of $m$. In
Fig.~\ref{fig:k3-Sigma} we plotted the complexity as a function of
$\alpha$, for various values of $m$. According to the interpretation
of the {\sf 1RSB} regime of Sec.~\ref{sec_reminder_rsb}, for each
value of $\alpha$ we can find the Parisi parameter $m_{\rm s}$ such that
$\Sigma=0$, and obtain the 1RSB estimate of the entropy as the
internal entropy of these states.
We plot this quantity in Fig.~\ref{fig:k3-phi_alpha},
together with the replica symmetric (RS) estimate and
the value obtained from the $m=0$ solution.

\begin{figure}
\includegraphics[width=0.6\textwidth]{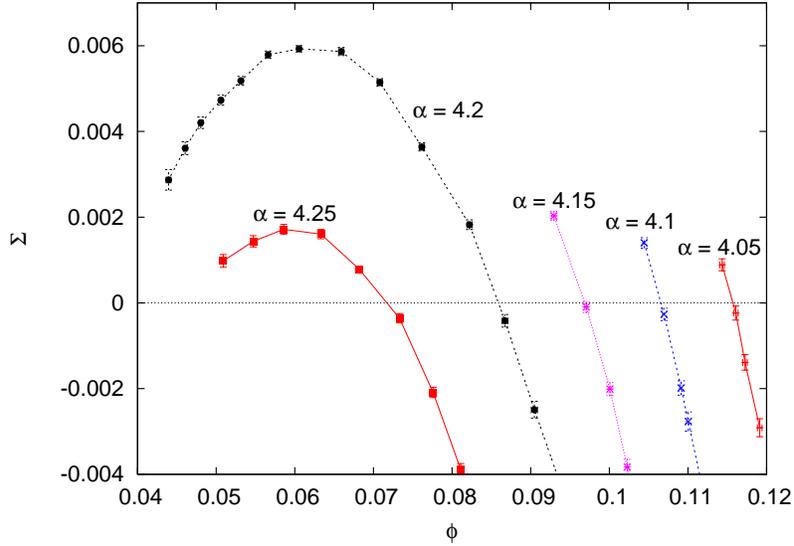}
\caption{The complexity $\Sigma(\phi)$ in random 3-SAT, for several values of
  $\alpha$.}
\label{fig:k3-Sigma_phi}
\end{figure}

We also present in Fig.~\ref{fig:k3-Sigma_phi} the entropic complexity
curves for a few values of $\alpha$.  Note that these curves can seem
incomplete; in fact for some values of $(\alpha,m)$ we found only
inconsistent solutions of the 1RSB equations, as is explained in more
details in Appendix~\ref{appendix_k3}. This might be related to an
instability of the 1RSB solution toward higher levels of replica
symmetry breaking \cite{MoRi,MoPaRi,FlLe2}.

\section{Simplifications of the 1RSB equations}
\label{sec_1rsb_tricks}

The numerical analysis of the 1RSB equations (\ref{eq_1RSB_Q}),
(\ref{eq_1RSB_P}) is, in general, an extremely difficult task.  Their
analytical control is even more challenging. In this section we
explain how the 1RSB approach simplifies in the two cases $m=0$ and
$m=1$, allowing for a precise numerical calculation of the complexity
and internal entropy in these points.

Because of the special role played by the value $m=1$, see Section
\ref{sec_reminder}, this enables to estimate precisely the dynamical
and condensation thresholds $\alpha_{\rm d}(k)$ and $\alpha_{\rm
  c}(k)$.  The simplifications arising at $m=0$ are on the other hand
the reason of the efficiency of the SP algorithm \cite{MeZe}.  Here we
will show how the states entropy can be computed at a small extra cost
with respect to the approach of \cite{MeZe}.

For the sake of concreteness, we discuss these simplification in the
case of random $k$-satisfiability. They have however a much wider
domain of validity. The same derivations do indeed hold for general
mean-field models on sparse random graphs.

%
%
\subsection{$m=1$ and  tree reconstruction}
\label{sec_1rsb_m1}

There is a strong connection between the 1RSB formalism with Parisi
parameter $m=1$ and the tree reconstruction problem (or computation of
point-to-set correlation), as discussed in~\cite{MeMo} and outlined in
Sec.~\ref{sec_reminder_sparse}. We follow here a somehow inverse
perspective with respect to~\cite{MeMo}: starting from the 1RSB
equations we shall progressively simplify them. At the end we shall
comment on their interpretation in terms of the tree reconstruction
problem.

Let us first define the averaging functional $\oh[P]$ (resp.\ $\ou[Q]$)
which associates to the distribution $P$ (resp.\ $Q$) of cavity fields
a single real through the relations
\begin{equation}
\tanh \oh[P] = \int \!\de P(h)\,  \tanh h \ , \qquad
\tanh \ou[Q] = \int \!\de Q(u)\, \tanh u \ .
\label{eq_def_oh}
\end{equation}
Consider now the right hand side of Eq.~(\ref{eq_1RSB_P}) for $m=1$.
The normalization factor can be expressed in terms of these averaged
fields,
\begin{equation}
\Z_3[Q_1^+,\dots,Q_{l_+}^+,Q_1^-,\dots,Q_{l_-}^-]  =
z_3[\ou[Q_1^+],\dots,\ou[Q_{l_+}^+],\ou[Q_1^-],\dots,\ou[Q_{l_-}^-]] \ .
\end{equation}
Using this fact and denoting by 
$G[Q_1^+,\dots,Q_{l_+}^+,Q_1^-,\dots,Q_{l_-}^-]$
the right hand side of Eq.~(\ref{eq_1RSB_P}) one can also show that
\begin{equation}
\oh[G[Q_1^+,\dots,Q_{l_+}^+,Q_1^-,\dots,Q_{l_-}^-] ] = 
\sum_{i=1}^{l_+} \ou[Q_i^+ ] - \sum_{i=1}^{l_-} \ou[Q_i^- ] \ .
\end{equation}
Treating similarly Eq.~(\ref{eq_1RSB_Q}), whose r.h.s. shall be denoted
$F[P_1,\dots,P_{k-1}]$, one obtains
\begin{equation}
\Z_4[P_1,\dots,P_{k-1}] = z_4(\oh[P_1],\dots,\oh[P_{k-1}]) \ , \qquad
\ou[F[P_1,\dots,P_{k-1}]] = f(\ou[P_1],\dots,\ou[P_{k-1}]) \ .
\end{equation}

$\oh$ (resp.\ $\ou$) can be viewed as a random variable, induced by
Eq.~(\ref{eq_def_oh}) with $P$ (resp.\ $Q$) drawn from $\P_{(1)}$
(resp.\ $\Q_{(1)}$). The above remarks show that their distributions
obey the RS self-consistency equation (\ref{eq_recurs_RS}).  Let us
now define a conditional average of $\P_{(1)}$, focusing on the $P$'s
in the support of $\P_{(1)}$ with a prescribed value of $\oh[P]$:
\begin{equation}
\oP(h|\oh) = \frac{1}{\P_{(0)}(\oh)} \int \!\de\P_{(1)}[P]\; \ P(h) \ 
\delta(\oh - \oh[P]) \ .
\label{eq_def_oP}
\end{equation}
The conditional distribution $\oQ(u | \ou )$ is defined analogously,
with $\P_{(1)}[P]$ replaced by $\Q_{(1)}[Q]$.

Consider again the distributional equations (\ref{eq_1RSB_Q}),
(\ref{eq_1RSB_P}).  Once the normalization factors have been expressed
in terms of the average fields $\oh$, $\ou$, the right-hand sides are
multi-linear functions of the distributions $P$, $Q$. It is thus
possible to take the conditional average as in Eq.~(\ref{eq_def_oP}).
This yields closed equations on $\oP$ and $\oQ$:
\begin{eqnarray}
\oQ(u|\ou)\Q_{(0)}(\ou) &=&
\int \prod_{i=1}^{k-1} \de\P_{(0)}(\oh_i) \; 
\delta(\ou-f(\oh_1,\dots,\oh_{k-1}))
\int \prod_{i=1}^{k-1} \;\de\oP(h_i|\oh_i) \ \delta(u-f(h_1,\dots,h_{k-1})) 
\frac{z_4(h_1,\dots,h_{k-1})}{z_4(\oh_1,\dots,\oh_{k-1})} \ ,
\nonumber\\
\oP(h|\oh)\P_{(0)}(\oh) &=& 
\sum_{l_+,l_-=0}^\infty \frac{e^{-\alpha k} (\alpha k/2)^{l_+ + l_-}}
{l_+ ! l_- !}
\int \prod_{i=1}^{l_+}\de\Q_{(0)}(\ou_i^+) 
\prod_{i=1}^{l_-}\de\Q_{(0)}(\ou_i^-)  
\ \delta\left(\oh- \sum_{i=1}^{l_+} \ou_i^+ + \sum_{i=1}^{l_-} \ou_i^- \right)
\nonumber \\
&& \int \prod_{i=1}^{l_+}\de\oQ(u_i^+|\ou_i^+) 
\prod_{i=1}^{l_-}\de\oQ(u_i^-|\ou_i^-)  
\ \delta\left(h-\sum_{i=1}^{l_+} u_i^+ + \sum_{i=1}^{l_-} u_i^- \right)
\frac{z_3(u_1^+,\dots,u_{l_+}^+,u_1^-,\dots,u_{l_-}^-)}
{z_3(\ou_1^+,\dots,\ou_{l_+}^+,\ou_1^-,\dots,\ou_{l_-}^-)} \ .
\label{eq_recurs_oP}
\end{eqnarray}
These equations are definitely simpler than the original ones
(\ref{eq_1RSB_Q}), (\ref{eq_1RSB_P}). In particular
$\oP(h|\oh)\P_{(0)}(\oh)$ can be viewed as a joint distribution of
$(h,\oh)$ and represented by a population of couples
$\{(h_i,\oh_i)\}_{i=1}^{\cal N}$. The presence of the reweighting
factors still represents a difficulty that we shall now get rid of by
a further simplification. Before proceeding, let us emphasize the
identities
\begin{equation}
\int \!\de\oP(h|\oh) \;\tanh h = \tanh \oh \ , \;\;\;\;\;\;\;\;\;\;\qquad 
\int \!\de\oQ(u|\ou)\; \tanh u = \tanh \ou \ ,
\label{eq_consistence}
\end{equation}
which follow directly from the definition (\ref{eq_def_oP}) and which
are indeed preserved by the equations (\ref{eq_recurs_oP}).  We define
now, for $\sigma =\pm 1$,
\begin{equation}
\oP_\sigma(h|\oh) = \frac{1+\sigma \tanh h}{1+\sigma \tanh \oh}\; \oP(h|\oh) \ .
\label{eq_def_oP_s}
\end{equation}
Using property (\ref{eq_consistence}), one can check that for any
$\oh$ and any $\sigma$ $\oP_\sigma(\bullet|\oh)$ is well normalized,
and that
\begin{equation}
\oP(h|\oh) = \sum_\sigma \frac{1+\sigma \tanh \oh}{2}\; \oP_\sigma(h|\oh) \ .
\label{eq_prop_oP}
\end{equation}
Similar definitions and properties hold for
$\oQ_\sigma(u|\ou)$. Inserting these definitions in
Eq.~(\ref{eq_recurs_oP}), one obtains
\begin{eqnarray}
\oQ_\sigma(u|\ou)\Q_{(0)}(\ou) &=&
\int \prod_{i=1}^{k-1} \!\de\P_{(0)}(\oh_i) \; 
\delta(\ou-f(\oh_1,\dots,\oh_{k-1})) \nonumber \\ && \hspace{5mm}
\sum_{\sigma_1,\dots,\sigma_{k-1}}
\mu(\sigma_1,\dots,\sigma_{k-1} | \sigma, \oh_1,\dots,\oh_{k-1} ) 
\int \prod_{i=1}^{k-1} \!\de\oP_{\sigma_i}(h_i|\oh_i) \; 
\delta(u-f(h_1,\dots,h_{k-1})) \ ,
\label{eq_recurs_oQ_s}
\end{eqnarray}
where the summation runs over the $2^{k-1}$ configurations of the
Ising spins $\sigma_1,\dots,\sigma_{k-1}$ with probabilities given by
\begin{eqnarray}
\mu(\sigma_1,\dots,\sigma_{k-1} | + , \oh_1,\dots,\oh_{k-1} ) &=&
\prod_{i=1}^{k-1} \frac{1+\sigma_i \tanh \oh_i}{2} \ , \label{eq_mu_p} \\
\mu(\sigma_1,\dots,\sigma_{k-1} | - , \oh_1,\dots,\oh_{k-1} ) &=&
\frac{\left(1- \I(\sigma_1 = \dots = \sigma_{k-1} = - )\right)}
{1- \prod_{i=1}^{k-1} \frac{1-\tanh \oh_i}{2}}
\prod_{i=1}^{k-1} \frac{1+\sigma_i \tanh \oh_i}{2} \ . \label{eq_mu_m}
\end{eqnarray}
The second of the equations in  (\ref{eq_recurs_oP}) yields
\begin{eqnarray}
\oP_\sigma(h|\oh)\P_{(0)}(\oh) &=& 
\sum_{l_+,l_-=0}^\infty \frac{e^{-\alpha k} (\alpha k/2)^{l_+ + l_-}}
{l_+ ! l_- !}
\int \prod_{i=1}^{l_+}\!\de \Q_{(0)}(\ou_i^+) 
\prod_{i=1}^{l_-}\!\de \Q_{(0)}(\ou_i^-)  
\ \delta\left(\oh- \sum_{i=1}^{l_+} \ou_i^+ + \sum_{i=1}^{l_-} \ou_i^-  \right)
\nonumber \\ && \hspace{15mm}
\int \prod_{i=1}^{l_+}\!\de \oQ_\sigma(u_i^+|\ou_i^+) 
\prod_{i=1}^{l_-}\!\de \oQ_{-\sigma}(u_i^-|\ou_i^-)  \ 
\delta\left(h- \sum_{i=1}^{l_+} u_i^+ + \sum_{i=1}^{l_-} u_i^-  \right) \ .
\label{eq_recurs_oP_s}
\end{eqnarray}

The equations (\ref{eq_recurs_oQ_s}), (\ref{eq_recurs_oP_s}) are
particularly convenient for numerical resolution.  This can be
obtained through an appropriate generalization of the population
dynamics algorithm, that employs two population of triples $\{(\oh_i,
h_i^+, h_i^-):\, i=1,\dots,\cN\}$ and $\{(\ou_j, u_j^+, u_j^-):\,
j=1,\dots,\cN\}$.  In the actual implementation it is actually more
convenient to store the hyperbolic tangent of these quantities,
e.g.\ $\tanh\oh_i$, $\tanh h_i^+$, etc.  These populations are updated
recursively according to the pseudocode below.  \vspace{0.5cm}

\begin{tabular}{ll}
\hline
\multicolumn{2}{l}{ {\sc Population Dynamics $m=1$} (Size $\cN$,
Iterations $t_{\rm max}$)}\\
\hline
1: & For all $i\in\{1,\dots,\cN\}$:\\
2: & \hspace{0.5cm}Set $h^{\pm}_i=\pm\infty$ and draw $\oh_i$ from 
$\P_{(0)}$;\\
3: & For all $t\in\{1,\dots,t_{\rm max}\}$:\\
4: & \hspace{0.5cm}For all $j\in\{1,\dots,\cN\}$ generate a new 
triple $(\ou_j, u_j^+, u_j^-)$:\\
5: & \hspace{0.85cm} Choose $k-1$ indices $i_1\dots i_{k-1}$ 
uniformly in  $[{\cal N}]$;\\
6: & \hspace{0.85cm} Compute $\ou_j = f(\oh_{i_1},\dots,\oh_{i_{k-1}})$;\\
7: & \hspace{0.85cm} Generate a configuration
$\sigma_1\dots\sigma_{k-1}$ with the law
$\mu(\cdots| +,\oh_{i_1}\dots\oh_{i_{k-1}})$
in Eq.~(\ref{eq_mu_p});\\
8: & \hspace{0.85cm} Compute 
$u_j^+ = f(h_{i_1}^{\sigma_1},\dots,h_{i_{k-1}}^{\sigma_{k-1}})$;\\ 
9: & \hspace{0.85cm} Generate a second configuration of spins with the 
law (\ref{eq_mu_m});\\
10: & \hspace{0.85cm} Set $u_j^- = f(h_{i_1}^{\sigma_1},\dots,h_{i_{k-1}}^{\sigma_{k-1}})$;\\
11: & \hspace{0.5cm} End-For;\\
12: & \hspace{0.5cm} For all $i\in\{1,\dots,\cN\}$ generate a new 
triple $(\oh_i, h_i^+, h_i^-)$:\\
13: & \hspace{0.85cm} Draw two independent
Poisson random variables $l_+$ and $l_-$ of mean $\alpha k/2$;\\
14: & \hspace{0.85cm} Draw $l_+ + l_- $ iid
indices $i_1^+,\dots,i_{l_+}^+,i_1^-,\dots,i_{l_-}^-$ uniformly random 
in $[{\cal N}]$;\\
15: & \hspace{0.85cm} Set
$\oh_j = \sum_{m=1}^{l_+} \ou_{i^+_m} - \sum_{m=1}^{l_-} \ou_{i^-_m}$,
 $h_j^\pm = \sum_{m=1}^{l_+} u^\pm_{i^+_m} - \sum_{m=1}^{l_-} u^\mp_{i^-_m}$;\\
16: & End-For;\\
\hline
\end{tabular}
\vspace{0.5cm}

The justification of the initialization will be given below.  After a
moment of thought one can convince oneself that the above update rules
are the correct discretization of Eqs.~(\ref{eq_recurs_oQ_s}) and
(\ref{eq_recurs_oP_s}).  More precisely, if the triples $(\oh_i,
h_i^+, h_i^-)$ are iid and the two pairs $(\oh_i, h_i^+)$, $(\oh_i,
h_i^-)$ have distributions (respectively) $\oP_+(h^+|\oh)
\P_{(0)}(\oh)$ and $\oP_-(h^-|\oh) \P_{(0)}(\oh)$, then the pairs
$(\ou_j, u_j^+)$, $(\ou_j, u_j^-)$ resulting from the above update
have distributions $\oQ_+(u^+|\ou) \Q_{(0)}(\ou)$, $\oQ_-(u^-|\ou)
\Q_{(0)}(\ou)$.  An analogous statement holds for the update from the
triples $(\ou_j, u_j^+, u_j^-)$ to $(\oh_i, h_i^+,
h_i^-)$\footnote{Notice that it would be wrong to claim that
  $(\oh_i,h_i^+,h_i^-)$ is distributed according to $\oP_+(h^+|\oh)
  \oP_-(h_-|\oh) \P_{(0)}(\oh)$~: the update rules used in the
  algorithm induce correlations between (for instance) the fields
  $h^+$ and $h^-$ inside the same triplet. These correlations do not
  spoil our claim.}.

Most relevant observables can be written as expectations with respect
to the distributions $\oP_\pm(h_\pm|\oh) \P_{(0)}(\oh)$,
$\oQ_\pm(u_\pm|\ou) \Q_{(0)}(\ou)$ and hence estimated from these
population of triplets.

Notice that, by definition, the 1RSB potential computed at $m=1$ is
equal to the RS free-entropy, $\Phi(m=1)=\phi_{(0)}$.  The internal
entropy can be expressed in terms of $\oP(h|\oh)$ and $\oQ(u|\ou)$ by
integrating over $\P_{(1)}$, $\Q_{(1)}$ in Eq.~(\ref{eq_phi_int}).
These conditional distributions can be further replaced by
$\oP_\sigma$ and $\oQ_\sigma$ thanks to Eq.~(\ref{eq_prop_oP}),
yielding finally
\begin{eqnarray}
\phi_{\rm int}(m=1) = &-& \alpha k \int \!\de \P_{(0)}(\oh) \de \Q_{(0)}(\ou)
\sum_\sigma \frac{1+\sigma \tanh(\ou + \oh)}{2} 
\int \!\de \oP_\sigma(h|\oh) \de \oQ_\sigma(u|\ou) \log z_1(u,h) \\
&+& \alpha \int \prod_{i=1}^k \!\de \P_{(0)}(\oh_i) 
\sum_{\sigma_1,\dots,\sigma_k}
\mu(\sigma_1,\dots,\sigma_k | \oh_1,\dots,\oh_k)
\int \prod_{i=1}^k \!\de \oP_{\sigma_i}(h_i | \oh_i) \log z_2(h_1,\dots,h_k) 
\nonumber \\
&+& \sum_{l_+,l_-=0}^\infty \frac{e^{-\alpha k} (\alpha k/2)^{l_+ + l_-}}
{l_+ ! l_- !} \int \prod_{i=1}^{l_+} \!\de \Q_{(0)}(\ou_i^+) 
\prod_{i=1}^{l_-} \!\de \Q_{(0)}(\ou_i^-) 
\sum_\sigma \frac{1 + \sigma \tanh\left(
\sum_{i=1}^{l_+} \ou_i^+ - \sum_{i=1}^{l_-} \ou_i^- \right)}{2}
\nonumber \\&& \hspace{2cm}
\int \prod_{i=1}^{l_+}\!\de \oQ_\sigma(u_i^+ | \ou_i^+)
\prod_{i=1}^{l_-} \!\de \oQ_{-\sigma}(u_i^- | \ou_i^-)
\log z_3(u_1^+,\dots,u_{l_+}^+,u_1^-,\dots,u_{l_-}^-) \ .
\nonumber
\end{eqnarray}
In the second term the distribution of the configuration
$(\sigma_1,\dots,\sigma_k)$ reads
\begin{equation}
\mu(\sigma_1,\dots,\sigma_k | \oh_1,\dots,\oh_k ) =
\frac{\left(1- \I(\sigma_1 = \dots = \sigma_k = - )\right)}
{1- \prod_{i=1}^k \frac{1-\tanh \oh_i}{2}}
\prod_{i=1}^k \frac{1+\sigma_i \tanh \oh_i}{2} \ .
\end{equation}
This expression of the internal free-entropy is readily evaluated by
sampling from the population of triplets defined above, the complexity
of the $m=1$ states is then finally expressed as $\Sigma(m=1)=
\Phi(m=1) - \phi_{\rm int}(m=1)$.

Consider now the definition of the overlaps given in
Eq.~(\ref{eq_def_overlaps}). The inter-state one $q_0$ is easily seen
to be equal to the RS one. Moreover $q_1$ can be written as
\begin{equation}
q_1 = \int \de\P_{(0)}(\oh) \int \de\oP(h|\oh) \tanh^2 h \ .
\end{equation}
To rewrite $q_1$ in terms of the distribution $\oP_\sigma$, note that 
$\tanh^2 h = (\tanh h) \sum_\sigma \sigma (1+\sigma \tanh h)/2 $ and use 
(\ref{eq_def_oP_s}) to obtain
\begin{equation}
q_1 = \int \de\P_{(0)}(\oh) \sum_\sigma \sigma \frac{1+\sigma \tanh \oh}{2}
\int \de\oP_\sigma(h|\oh) \tanh h \ .
\end{equation}
These expressions allow to estimate $q_0$, $q_1$ from the population
of triples $\{(\oh_i,h^+_i,h^-_i)\}$.

In Figs.~\ref{fig:k4_C_ell} and \ref{fig:k3_C_ell} we followed this
approach to plot the difference $q_1(\ell)-q_0$ for several values of
$\alpha$ and $k=3,4$, whereby the population $\{(\oh_i,h^+_i,h^-_i)\}$
is obtained after $\ell$ iterations of the above algorithm.  For
$\alpha<\alpha_{\rm d}(k)$, $q_1(\ell)-q_0\stackrel{\ell}{\to} 0$,
while for $\alpha>\alpha_{\rm d}(k)$ it is bounded away from $0$.  Let
us emphasize the great simplification achieved: the equations
(\ref{eq_recurs_oQ_s},\ref{eq_recurs_oP_s}) are much simpler than the
original 1RSB equations: they can be solved using a simple population
of triples, instead of a population of populations. Further, the
initialization used in the pseudocode above is the \emph{correct} one,
in the following sense.  If the equations (\ref{eq_recurs_oQ_s}),
(\ref{eq_recurs_oP_s}) admit a non-trivial solution, then their
iteration converges to a non trivial solution under such an
initialization.

The last statement follows from the interpretation of the order
parameters in terms of tree reconstruction.  Consider an infinite tree
$k$-satisfiability formula roted at variable node $i$. The tree is
random with distribution defined by letting each variable to be
directly (resp. negated) in $l_+$ (resp. $l_-$) clauses, where
$l_{\pm}$ are independent random Poisson random variables with mean
$\alpha k/2$. One can define a (uniform) free boundary Gibbs measure
$\mu$ over SAT assignments of such a tree.  Imagine now to generate a
solution from this measure, conditional on the root value being
$\sigma$, and denote by $\us_B$ the values of variables at distance at
least $\ell$ from the root. Define the fields $\oh$,
$h^{\sigma}_{\ell}$ by
\begin{eqnarray}
\mu(\sigma_i) \equiv \frac{1+\sigma_i\tanh \oh}{2}\, ,\;\;\;\;\;\;
\mu(\sigma_i|\us_B) \equiv \frac{1+\sigma_i\tanh h^{\sigma}_{\ell}}{2}\, .
\end{eqnarray}
Notice that both are random quantities, $\oh$ because of the tree
randomness and $h^{\pm}_{\ell}$ both because of the tree and of the
random configuration $\us_B$.  Let $\oP_{\sigma}^{\ell}(h|\oh)$ be the
conditional distribution of $h^{\sigma}_{\ell}$ given $\oh$.

It is not hard to show that $\oP_{\sigma}^{\ell}(h|\oh)$ is the
distribution obtained by iterating (\ref{eq_recurs_oQ_s}),
(\ref{eq_recurs_oP_s}) $\ell$ times with initial condition
$\oP_{\pm}^{\ell}(h|\oh) = \P_{(0)}(\oh)\delta(h \mp \infty)$.  This
corresponds indeed to the initialization we used in the population
dynamics algorithm. It follows from the arguments in \cite{MeMo} that
this is the \emph{correct} initialization, in the sense described
above. Further, under the usual assumptions of the cavity method and
for $\alpha<\alpha_{\rm c}(k)$, the quantity $q_1(\ell)-q_0$ plotted
in Figs.~\ref{fig:k4_C_ell} and \ref{fig:k3_C_ell} coincides with the
correlation function (\ref{eq:Correlation}) in the large $N$ limit.
%
%
\subsection{$m=0$: Survey Propagation and the associated internal entropy}
\label{sec_1rsb_m0}

We turn now to the second particular case for which a simplified
treatment of the 1RSB formalism is possible, namely at $m=0$.

To begin with, let us consider the structure of the distributions
$P(h)$ (resp.\ $Q(u)$) in the support of $\P_{(1)}$ (resp.\ $\Q_{(1)}$)
for an arbitrary value of $m$. A moment of thought reveals the
possibility of ``hard fields'' $h=\pm \infty$ that strictly constrains
a variable to take the same value in all configurations of a cluster
of solutions. We can take care explicitly of this possibility by
denoting
\begin{equation}
P(h) = x^- \delta(h+\infty) + x^+ \delta(h-\infty) 
+ (1- x^- -x^+) \tP(h) \ , \qquad 
Q(u) = y \ \delta(u-\infty) + (1-y) \tQ(u) \ , 
\label{eq_def_hard}
\end{equation}
where $\tP$ and $\tQ$ have their support on finite values of the
fields, that shall be called `soft' or `evanescent'.  Rewriting the
right hand side of (\ref{eq_1RSB_Q}) with these notations yields
\begin{eqnarray}
Q(\bullet) &\eqd& \frac{1}{\Z_4[P_1,\dots,P_{k-1}]} \left[  
\left(\prod_{i=1}^{k-1} x^-_i \right) \delta(\bullet-\infty) 
+  2^m \left( 1 - \prod_{i=1}^{k-1}(1-x^+_i)\right) \delta(\bullet) 
\right. \nonumber \\ && \left. 
+ \sum_{|I|\ge 1} \prod_{i \in I} (1-x^+_i-x^-_i)
\prod_{i\notin I} x^-_i \int \prod_{i \in I} \de\tP_i(h_i) (1+e^{-2\bullet})^m
\delta\left(\bullet+\frac{1}{2} 
\log \left(1 - \prod_{i \in I} \frac{1 -\tanh h_i}{2} \right) \right) 
\right]
\label{eq_1RSB_Q_T0}
\end{eqnarray}
where the summation on $I$ is over the non empty subsets of
$\{1,\dots,k-1\}$.

To achieve the same task for Eq.~(\ref{eq_1RSB_P}) it is advisable to
introduce some more compact notations,
\begin{equation}
\pi_\sigma = \prod_{i=1}^{l_\sigma} (1-y_i^\sigma) \ , \qquad
S_\sigma = \prod_{i=1}^{l_\sigma} (1+ \tanh u_i^\sigma) \ , \qquad
T_\sigma = \prod_{i=1}^{l_\sigma} (1 - \tanh u_i^\sigma) \ ,
\label{eq_shorthand}
\end{equation}
in terms of which we have for instance $z_3 = S_+ T_- + T_+ S_-$.  We
shall also denote $\e[\bullet]$ the average over the $u_i^{\pm}$ drawn
from the $Q_i^{\pm}$, and $\te$ similarly using $\tQ_i^{\pm}$.  We
then obtain
\begin{eqnarray}
P(\bullet) &\eqd& \frac{1}{\Z_3[\{Q_i^+\},\{Q_i^-\}]} \left[
\pi_+ \te[T_+^m] \left(\e[S_-^m] - \pi_- \te[S_-^m]\right) 
\delta(\bullet+\infty)
\ + \ \pi_- \te[T_-^m] \left(\e[S_+^m] - \pi_+ \te[S_+^m]\right) 
\delta(\bullet-\infty)
\phantom{\prod_{i=1}^{l_-}}
\right. \nonumber \\ && \left.
\hspace{2cm}
+ \ \pi_+ \pi_- 
\int \prod_{i=1}^{l_+} \de\tQ_i^+(u_i^+) \prod_{i=1}^{l_-} 
\de\tQ_i^-(u_i^-) \ 
\delta\left(\bullet - \sum_{i=1}^{l_+} u_i^+ + \sum_{i=1}^{l_-} u_i^- \right)
(S_+ T_- + T_+ S_-)^m
\right]
\label{eq_1RSB_P_T0}
\end{eqnarray}

Analogously, the replicated free-entropy $\Phi(m)$ and its derivative
can be rewritten by making explicit the distinction between hard and
soft fields.

Consider now the previous equations with $m=0$. As we have explicitly
removed all the contradictory terms which had a strictly vanishing
reweighting factor in the original
relations~(\ref{eq_1RSB_Q},\ref{eq_1RSB_P}), all the terms raised to
the power $m$ in Eqs.~(\ref{eq_1RSB_Q_T0},\ref{eq_1RSB_P_T0}) are
strictly positive, hence these factors go to 1 when $m$ vanishes. Two
important consequences are to be underlined~: the normalization
factors $\Z_3$ and $\Z_4$ do not depend on the evanescent
distributions $\tP$, $\tQ$. In fact $\Z_3 = \pi_+ + \pi_- - \pi_+
\pi_-$ and $\Z_4=1$. Moreover the equations on the intensity of the
hard fields peaks decouple from the evanescent part when $m$ goes to
0, (\ref{eq_1RSB_Q_T0},\ref{eq_1RSB_P_T0}) yielding for them
\begin{equation}
y \eqd \prod_{i=1}^{k-1} x^-_i \ , \qquad 
(x^+,x^-) \eqd \left( 
\frac{(1-\pi_+) \pi_- }{\pi_+ + \pi_- - \pi_+ \pi_-},
\frac{(1-\pi_-) \pi_+ }{\pi_+ + \pi_- - \pi_+ \pi_-} \right) \ ,
\end{equation}
which are nothing but the probabilistic form of the Survey Propagation
equations~\cite{MeMeZe}. For future use we denote $Q_{\rm SP}(y)$ and
$P_{\rm SP}(x_+,x_-)$ the distributions of these random variables.
The complexity at $m=0$ is $\Sigma(m=0)=\Phi(m=0)$ and can then be
expressed from Eq.~(\ref{eq_1RSB_Phi}) as
\begin{equation}
\Sigma(m=0)=\Phi(m=0)=-\alpha k \E [\log (1 -x^- y) ] + 
\alpha \E \left[ \log \left(1 - \prod_{i=1}^k x^-_i \right) \right]
+ \E [\log ( \pi_+ + \pi_- - \pi_+ \pi_-  )] \ ,
\end{equation}
where the average is done with respect to $P_{\rm SP}$ and $Q_{\rm SP}$.

By focusing on the intensity of the hard fields this 'energetic'
version of the cavity method~\cite{MeZe,MeMeZe} lost the information
contained in the evanescent field distributions $\tP$, $\tQ$, which is
necessary to obtain the internal entropy of the states,
$\Phi'(m=0)$. This quantity can however be obtained in a rather simple
way. We shall indeed define $\tQ(u|y)$ as the average of the
evanescent part of $Q$ drawn from $\Q_{(1)}$, conditioned on the value
of the hard field delta peak, and similarly $\tP(h|x^+,x^-)$. As the
right hand sides of (\ref{eq_1RSB_Q_T0},\ref{eq_1RSB_P_T0}) are linear
functionals of these evanescent distributions when $m=0$, closed
equations on this conditional averages can be obtained. We shall write
them in terms of the joint distributions $\tQ(u,y) = \tQ(u|y) Q_{\rm
  SP}(y)$ and $\tP(h,x^+,x^-) = \tP(h|x^+,x^-) P_{\rm SP}(x^+,x^-)$,
\begin{multline}
\tQ(u,y) = \int \prod_{i=1}^{k-1} \de h_i \de
x^+_i \de x^-_i \tP(h_i,x^+_i,x^-_i)
\ \delta\left(y- \prod_{i=1}^{k-1} x^-_i \right) \left[
\frac{1 - \prod_{i=1}^{k-1} (1-x^+_i) }{1-y} \ \delta(u) \right. 
\\ 
\left. +
\frac{\sum_{p=1}^{k-1} \binom{k-1}{p}
\prod_{i=1}^p (1-x^+_i-x^-_i) \prod_{i=p+1}^{k-1} x^-_i }{1-y}
\delta\left( u + \frac{1}{2} \log\left(1 - \prod_{i=1}^p \frac{1-\tanh h_i}{2} 
\right)  \right)
\right] \ ,
\label{eq_tQ}
\end{multline}

\begin{multline}
\tP(h,x^+,x^-) = \sum_{l_+,l_-=0}^\infty 
\frac{e^{-\alpha k} (\alpha k/2)^{l_+ + l_-}}{l_+! l_-!} \int
\prod_{i=1}^{l_+} \de u_i^+ \de y_i^+ \tQ(u_i^+,y_i^+) 
\prod_{i=1}^{l_-} \de u_i^- \de y_i^- \tQ(u_i^-,y_i^-) \\
\delta \left( 
x^+ -  \frac{(1-\pi_+) \pi_- }{\pi_+ + \pi_- - \pi_+ \pi_-}\right)
\ \delta \left( 
x^- -  \frac{(1-\pi_-) \pi_+ }{\pi_+ + \pi_- - \pi_+ \pi_-}\right)
\ \delta\left( h - \sum_{i=1}^{l_+} u_i^+ + \sum_{i=1}^{l_-} u_i^- \right)
\ .
\label{eq_tP}
\end{multline}

A solution of these equations can be obtained through a simple
population dynamics algorithm, encoding $\tQ(u,y)$ as a population of
couples $\{(u_i,y_i) \}_{i=1}^{\cal N}$ and $\tP(h,x^+,x^-)$ as
$\{(h_i,x^+_i,x^-_i)\}_{i=1}^{\cal N}$. The update rules of the
algorithm can be deduced from (\ref{eq_tQ},\ref{eq_tP}): a new element
$(h,x^+,x^-)$ is obtained drawing two Poisson random variables $l_\pm$
of mean $\alpha k/2$, $l_+ + l_-$ elements of the population
$\{(u_i,y_i) \}$ and combining them according to (\ref{eq_tP}). The
translation of (\ref{eq_tQ}) is only slightly more complicated. After
extracting $k-1$ elements at random from the population
$\{(h_i,x^+_i,x^-_i)\}$ one obtains $y$ as the product of the $k-1$
elements $x_-$. One then draws a configuration $(s_1,\dots,s_{k-1})\in
\{-1,0,+1 \}^{k-1}$, each `spin' $s_i$ being $\pm 1$ with probability
$x_i^\pm$ and 0 with probability $1 - x_i^+ - x_i^-$, conditional on
$(s_1,\dots,s_{k-1})\neq(-1,\dots,-1)$.  If at least one of the $s_i$
is equal to $+1$ the new value of $u$ is taken to $0$, otherwise
$u=-\log(1 - \prod(1-\tanh h_i)/2)/2$, the product being taken on the
indices $i$ such that $\s_i=0$.

The internal entropy of the $m=0$ pure states can be obtained from the
solution of these equations, simplifying Eq.~(\ref{eq_phi_int}) into
\begin{eqnarray}
\phi_{\rm int}(m=0) = &-&\alpha k \E \left[ \frac{
x^+ y \log 2 + (1-y) (x^- \log(1-\tanh u) + x^+ \log(1+\tanh u))
}{1-x^- y}
\right] \\
&-&\alpha k \E \left[ \frac{
(1-x^+-x^-) (y  \log(1+\tanh h) + (1-y)  \log(1+\tanh h \tanh u)) 
}{1-x^- y} 
\right] \\
&+& \alpha \E \left[\sum_{p=1}^k \binom{k}{p}
\prod_{i=1}^p (1-x^+_i-x^-_i) \prod_{i=p+1}^k x^-_i \log\left( 
1 - \prod_{i=1}^p \frac{1-\tanh h_i}{2}\right)
\right]\\
&+& \E \left[ \pi_+ \pi_- \log(S_+ T_- + T_+ S_- ) 
+ \pi_- (1-\pi_+) \log(S_+ T_- ) + \pi_+ (1- \pi_-) \log(T_+ S_-)  \right]
\ ,
\end{eqnarray}
where the expectation is over independent copies of elements drawn
from $\tP(h,x^+,x^-)$ and $\tQ(u,y)$, and in the last line (where we
used the shorthand notations defined in (\ref{eq_shorthand})) over the
Poissonian random variables $l_\pm$. This quantity was plotted for
$k=4$ in Fig.~\ref{fig:k4-Sigma}.

Let us emphasize the great numerical simplification with respect to
the general 1RSB equations: we have to deal here with populations of
couples (or triplet) of fields, not populations of populations. Yet we
manage to extract not only the complexity, which was the one computed
in the probabilistic version of survey propagation, but also the
associated internal entropy.

\section{Large $k$ results}
\label{sec_large_k}

To complement the numerical resolution of the 1RSB equations, we
present in this Section analytic expansions of the various thresholds
and thermodynamic quantities for large $k$. Some technical details of
these computations are deferred to Appendix \ref{app_largek}.
%
%
\subsection{Dynamical transition regime}
\label{sec:DynamicalTr}

A non-trivial solution of the 1RSB equations appears in the regime
defined by
\begin{eqnarray}
\alpha_{\rm d} = \frac{2^k}{k}\left[\log k + \log\log k + \gamma
+ O\left(\frac{\log \log k}{\log k}\right) \right]\, ,
\end{eqnarray}
with $\gamma$ finite as $k\to\infty$.  In this regime the 1RSB
distributional order parameters $\P_{(1)}$, $\Q_{(1)}$ are supported
on cavity field distributions of the form (\ref{eq_def_hard}) with
$\tP(\,\cdot\,)$, $\tQ(\,\cdot\,)$ supported on finite fields. The
weights of the hard fields are deterministic to leading order, with
\begin{eqnarray}
x^{\pm}  = 
\frac{1}{2}-\frac{\delta(\gamma,m)}{2k\log k} + 
O\left(\frac{1}{k(\log k)^2}\right)\, ,\;\;\;\;\;\;\;\;\;\;\;
y  =\frac{1}{2^k}\, 2^{1-m}\left\{1-\frac{\delta(\gamma,m)}{\log k}+
O\left(\frac{1}{(\log k)^2}\right)\right\}\, .
\end{eqnarray}
A set of coupled equations can also be written for the averages of
$\tP$, $\tQ$, in terms of which one computes a function
$\Lambda(\delta,m)$ that finally determines $\delta(\gamma,m)$ as a
function of $\gamma$ by solving the following equation:
\begin{eqnarray}
\gamma = \delta+\log \frac{1}{2\delta} +\Lambda(\delta,m)\, .
\label{eq:gamma}
\end{eqnarray}
Both the expressions for $\Lambda(\delta,m)$ and the equations for the
averages of $\tP$, $\tQ$ are quite involved and we report them in
Appendix \ref{app_largek}. In any case the right hand side of
Eq.(\ref{eq:gamma}) diverges for $\delta\to 0$ and
$\delta\to\infty$. As a consequence a pair of
solutions\footnote{Consistency arguments imply that the one with
  smaller $\delta$ must be selected.}  appears for
$\gamma\ge\gamma_{\rm d}(m)$, where $\gamma_{\rm d}(m)$ is obtained by
minimizing the above expression over $\delta$.  For $m=0,1$ the
formulae simplify yielding $\Lambda(\delta,m=0) = 0$ and
$\Lambda(\delta,m=1) =\log 2$ independently of $\delta$, whence the
minimum takes place at $\delta = 1$ for these two values of $m$.

To summarize this yields the following estimate for the dynamical
threshold
\begin{equation}
\alpha_{\rm d}(k,m)= \frac{2^k}{k}\left[\log k + \log\log k + \gamma_{\rm d}(m)
+ O\left(\frac{\log \log k}{\log k}\right) \right] \ ,
\end{equation}
with $\gamma_{\rm d}(m=1)=1$ and $\gamma_{\rm d}(m=0)=1-\log 2$.
Notice that the transition at $m=1$ occurs slightly after the one at
$m=0$ in agreement with what is found numerically for small values of
$k \ge 4$.
%
%
\subsection{Intermediate regime}

Consider now the limit $k \to \infty$ with $\alpha = 2^k \halpha$ for
some fixed $\halpha >0$. On this scale the SAT/UNSAT phase transition
occurs at $\halpha_{\rm s} = \log 2+O(2^{-k})$~\cite{MeMeZe,largek}.
We shall therefore assume $\halpha\in (0,\log 2)$. In this regime it
is convenient to use again the decomposition (\ref{eq_def_hard}), with
at leading order $\tP(h) = \delta(h)$ and $x^{\pm} = (1/2) (1-\hx^\pm
e^{-\halpha k})$. From this Ansatz one finds that $\hx^{\pm} =
2^{m-1}$, then it follows that the 1RSB potential is asymptotically
\begin{equation}
\Phi(m) = \log 2 - \halpha + e^{-\halpha k} (2^{m-1} -1) 
+ O(2^{-k})\ .
\label{eq_res_intermediate}
\end{equation}
By derivation of this expression one obtains the internal entropy,
\begin{equation}
\phi_{\rm int}(m) = e^{-\halpha k} 2^{m-1} \log 2 + O(2^{-k}) \ , \qquad
\end{equation}
and defining a reduced quantity $\s$ by $\phi_{\rm int}=e^{-\halpha k}
(\log 2) \s$, we get the complexity function explicitly,
\begin{equation}
\Sigma(\s) = \log 2 - \halpha + e^{-\halpha k}  \tSigma(\s) + O(2^{-k}) \ , 
\qquad
\tSigma(\s) = \s (1-\log 2) - \s \log \s -1 \ .
\end{equation}
Notice that, for large $k$, the internal entropy of states is
exponentially smaller (in $k$) than the complexity.  Further, to
leading order, the complexity vanishes at $\halpha =\log 2$,
independently on $m$.
%
%
\subsection{Condensation regime}

In order to resolve the separation between the condensation and
satisfiability phase transitions we must let $k\to\infty$ with
$\alpha\simeq 2^k\log 2$. More precisely, we define $\alpha = 2^k \log
2 - \zeta$, and take $k\to\infty$ with $\zeta$ fixed. Again, we use
the Ansatz (\ref{eq_def_hard}) with, at leading order $\tP(h) =
\delta(h)$ and $x^{\pm} = (1/2) (1-\hx^\pm 2^{- k})$.

We then get the expansion of the potential,
\begin{equation}
%
\Phi(m) = \frac{1}{2^{k}}\, \{\zeta-\zeta_{\rm s}+(2^{m} -1)/2\}
+ O(2^{-2k})\, ,
\end{equation}
with $\zeta_{\rm s} \equiv\frac{1}{2}(1+\log 2)$. The entropy can be
determined by deriving the above with respect to $m$; defining the
reduced entropy density through $\phi_{\rm int}=2^{-k} (\log 2) \s$,
the complexity reads in this regime
\begin{equation}
\Sigma(\s) = \frac{1}{2^k} \left\{\zeta-\zeta_{\rm s}+ \s (1-\log 2) -
  \s \log \s -\frac{1}{2}\right\} +O(2^{-2k})\, .
\end{equation}
The condensation and satisfiability transition are located by
determining $\zeta$ such that $\Sigma(m)=0$ for (respectively)
$m=1$ and $m=0$. We get
\begin{equation}
\alpha_{\rm c}(k) = 2^k \log 2 - \frac{3 \log 2}{2} + O(2^{-k}) \ , \qquad
\alpha_{\rm s}(k) = 2^k \log 2 - \frac{1 + \log 2}{2} + O(2^{-k}) \ .
\end{equation}
The thermodynamic value $m_{\rm s}(\zeta)$ of the Parisi parameter between
these two thresholds is obtained by minimizing $\Phi(m)/m$. At the
order of the expression of $\Phi(m)$ given above $m_{\rm s}(\zeta)$ is
solution of
\begin{eqnarray}
\zeta -\zeta_{\rm s} = 2^{m-1}(2^{-m}-1+m\log 2)\, .
\label{eq_m_largek}
\end{eqnarray}
In particular one finds close to the satisfiability transition
\begin{equation}
m_{\rm s}(\zeta) \simeq \frac{2}{\log 2} \sqrt{\zeta - \zeta_{\rm s}} \ .
\label{eq_m_largek2}
\end{equation}

\begin{figure}
\includegraphics[width=0.45\textwidth]{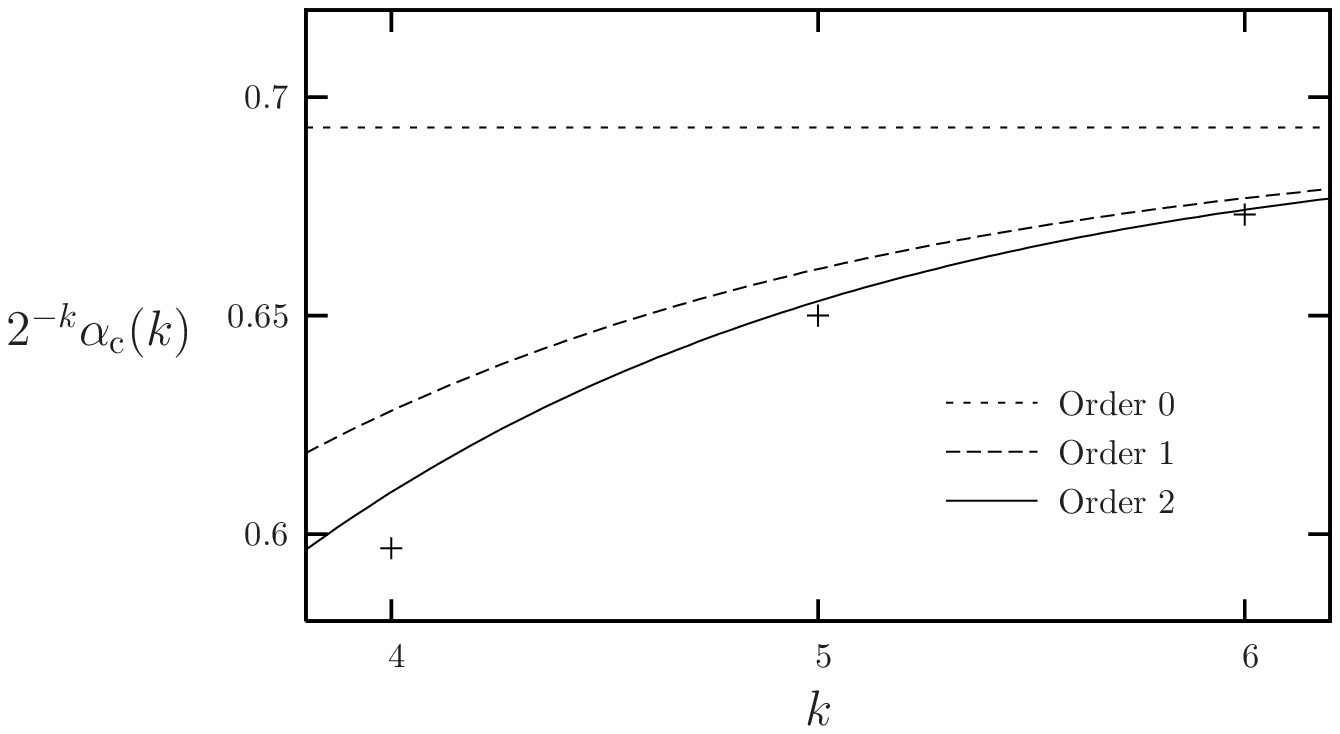}
\caption{Condensation threshold in reduced units, $2^{-k}\alpha_{\rm
    c}(k)$.  Symbols: numerical determination by population dynamics
  algorithm, see Tab.~\ref{tab_thresholds}.  Lines: analytical large
  $k$ expansion, truncated at the three first orders, see
  Eq.~(\ref{eq_condensation_res}).}
\label{fig_expansion}
\end{figure}

A systematic expansion in powers of $2^{-k}$ of the satisfiability
threshold $\alpha_{\rm s}(k)$ has been performed up to seventh order
in \cite{MeMeZe}.  The corresponding expansion for the condensation
threshold $\alpha_{\rm c}(k)$ is slightly more difficult, because of
the necessary control of the corrections to the evanescent field
distributions. We thus contented ourselves with the computation of the
next order in the expansion,
\begin{eqnarray}
\alpha_{\rm c}(k) = 2^{k}\log 2 &-& \frac{3 \log 2 }{2}  
\label{eq_condensation_res}\\ &-& 
\left[ \frac{6 (\log 2)(\log 3) - 7 (\log 2)^2 }{4} k^2 
+ \frac{5 (\log 2)^2 - 3 (\log 2)(\log 3)}{2} k - \frac{5 \log 2}{12}
\right]\frac{1}{2^{k}} + O\left({\rm poly}(k)\frac{1}{2^{2k}} \right) \ .
\nonumber
\end{eqnarray}
This expression is compared in Fig.~\ref{fig_expansion} with the
numerical results for small $k$.

\section{Conclusion}
\label{sec_conclusions}

The set of solutions of random $k$-satisfiability formulae exhibits a
surprisingly rich structure, that has been explored in a series of
statistical mechanics studies \cite{MoZe,BiMoWe,MeZe}.  Either
implicitly or explicitly, these studies are based on defining a
probability distribution over the solutions, and then analyzing its
properties. While the most natural choice is the uniform measure, the
authors of Ref.~\cite{MeZe} achieved a great simplification (and a
wealth of exact results) by implicitly weighting each solution
inversely to the size of the `cluster' it belongs to.  Since clusters
sizes are exponential in the number of variables, and have large
deviations, this amounts to focusing on an exponentially small subset
of solutions.

In this paper we resumed the (technically more challenging) task of
studying the uniform measure and obtained the first complete phase
diagram (including replica symmetry breaking) in this setting.  While
we confirmed several of the predictions in \cite{MeZe}, our analysis
unveiled a number of new phenomena:
\begin{enumerate}
\item There exists a critical value $\alpha_{\rm d}(k)$ of the clause
  density that can be characterized in several equivalent ways: $(i)$
  Divergence of auto-correlation time under Glauber dynamics; $(ii)$
  Divergence of point-to-set correlation length; $(iii)$ Appearance of
  bottlenecks between `sizable' subsets of solutions.  The value of
  $\alpha_{\rm d}(k)$ is bigger than the value obtained with the
  method of~\cite{MeZe} (except for $k=3$ where it is smaller).
\item While $\alpha_{\rm d}(k)$ does not correspond to an actual
  thermodynamic phase transition, such a phase transition takes place
  at a second threshold $\alpha_{\rm c}(k)<\alpha_{\rm s}(k)$
  ($\alpha_{\rm s}(k)$ being the satisfiability threshold).  This
  manifests in two-point correlations, as well as in the overlap
  distribution.
\item The phase diagram is qualitatively different for $k \ge 4$ and
  $k=3$. The latter value has been most commonly used in numerical
  simulations. This difference had not been recognized before because
  it does not show up in the behavior of the maximal complexity
  $\Sigma(m=0)$ investigated up to now.
\end{enumerate}

A number of research directions are suggested by this refined
understanding:
\begin{enumerate}
\item[$(a)$] We kept ourselves to 1RSB: it would be extremely
  interesting to investigate whether more complex hierarchical (FRSB)
  structures can arise in the set of solutions. A first step in this
  direction would be to analyze the stability~\cite{MoPaRi,MoRi,FlLe2}
  of the 1RSB Ansatz, in particular to clarify our numerical findings
  for $k=3$. For $k\ge 4$ we believe that our determination of
  $\alpha_{\rm d}$ and $\alpha_{\rm c}$ is not affected by FRSB, yet
  it might be that the pure states, for some values of their internal
  entropy, are to be described by a FRSB structure.
\item[$(b)$] The dynamical threshold $\alpha_{\rm d}(k)$ is expected
  to affect algorithms that satisfy detailed balance with respect to
  the uniform measure over solutions (or its positive temperature
  version). Let us stress that it is likely not to have any relation
  with more general local search
  algorithms~\cite{LocalSearch1,LocalSearch2,LocalSearch3}.  It is an
  open problem to generalize the static computations performed here to
  obtain meaningful predictions in those cases.
\item[$(c)$] Finally, the discovery of the condensation phase
  transition at $\alpha_{\rm c}(k)$ suggests that belief propagation
  might be effective in computing marginals up to this threshold, as
  the average of the 1RSB equations with $m=1$ corresponds to BP.  The
  possible use of this information in constructing solutions is
  discussed in~\cite{Maneva,Pretti,Allerton,decimation_future}.
\end{enumerate}

\section*{Acknowledgments}

We thank Florent Krzakala and Lenka Zdeborova for several discussions on this
project.
%
%
\appendix

\section{On the numerical resolution of the 1RSB cavity equations}
\label{sec_numerics}

In this section we discuss some issues related to the numerical
resolution of Eqs.~(\ref{eq_1RSB_Q}), (\ref{eq_1RSB_P}).  As already
mentioned, the 1RSB order parameter $\P_{(1)}[P]$ is approximated by a
sample of ${\cal N}$ populations, each composed of ${\cal N}'$
elements $h_{i,j}$, $i\in [\cN]$, $j\in [\cN']$.  The numerical
results presented in this work have been obtained with ${\cal N}=10^4$
and ${\cal N}'=10^3$.

The solution to the 1RSB cavity equations is found by an iterative
procedure: starting from a ``good'' initial guess for the fixed point
solution, we iterate a sampled version of Eqs.~(\ref{eq_1RSB_Q}),
(\ref{eq_1RSB_P}).  After some iterations the sample of populations
converges to a stationary state with fluctuations of order
$O(1/\sqrt{{\cal N}},1/\sqrt{{\cal N}'})$.  Convergence to the
stationary regime is usually fast and may take around $10^2$
iterations in the worst cases we encountered.  Once in the stationary
regime, we keep iterating for at least $10^4$ steps. Meanwhile we take
averages (over the populations and over the time evolution) of the
quantities of interest.  This considerably reduces statistical errors.

Our actual numerical implementation makes use of two transformations
with respect to Eqs.~(\ref{eq_1RSB_Q}), (\ref{eq_1RSB_P}).  First, we
make a change of variables into
\begin{equation}
\varphi = e^{-2 u}\;, \qquad \psi = \frac{1+\tanh(h)}{2}\;,
\end{equation}
both taking values in $[0,1]$ (note that the variable $u$ is defined
non-negative, see the definition of the function
$f(h_1,\ldots,h_{k-1})$ in Eq.(\ref{eq_recurs_def1})).  Moreover we
exploit the fact that the reweighting term $z_4(h_1,\dots,h_{k-1})$ in
Eq.~(\ref{eq_1RSB_Q}) is a function of $u=f(h_1,\dots,h_{k-1})$
(cf. Eq.~(\ref{eq_def_z4})). This allows to transfer all the effects
of reweighting to the other equation.  Denoting $\hQ(\varphi)$ and
$\hP(\psi)$ the new distributions, these two transformations lead to
\begin{eqnarray}
\hQ(\bullet) &\eqd& \int \prod_{i=1}^{k-1}\!\de \hP_i(\psi_i)\;
\delta\left[\bullet - 1 + \prod_i(1-\psi_i)\right]\label{eq:hQ}\\
\hP(\bullet) &\eqd& \frac{1}{Z} 
\int \prod_{i=1}^{l_+} \!\de \hQ_i^+(\varphi_i^+)
\prod_{i=1}^{l_-} \!\de \hQ_i^-(\varphi_i^-)\;
\delta\left[\bullet - \frac{\prod_i
    \varphi_i^-}{\prod_i \varphi_i^+ + \prod_i \varphi_i^-}\right]
\left(\prod_i \varphi_i^+ + \prod_i \varphi_i^-\right)^m\label{eq:hP}
\end{eqnarray}
where $Z$ in the last equation is obtained by normalization.

One delicate issue in solving this kind of equation is how to
represent faithfully the left hand side of Eq.~(\ref{eq:hP}) by a
sample of ${\cal N}'$ representative elements of $\hP$, because of the
reweighting term $(\prod_i \varphi_i^+ + \prod_i \varphi_i^-)^m$.  A
possible solution~\cite{MePa} consists in first generating a larger
number, say $5{\cal N}'$, of outgoing fields, store them along with
the associated weights, and then perform a resampling step to extract
${\cal N}'$ elements from this intermediate population.  This approach has the
advantage of having complexity independent of the distributions
$\hQ_i$. Unhappily if the weights are strongly concentrated on a small
subset of the $5\cN'$ fields, the resampled population will have many
copies of these elements. This leads to a deterioration of the sample.

We adopted a different strategy whose running time depends on how
strong is the reweighting.  For $m \ge 0$, we generate fields
sequentially, and include them in the new population with probability
proportional to the reweighting factor (divided by the
normalization factor $2^m$).  This procedure becomes slower when $m$
grows, but it ensures that no repetitions appear in the new sample.
 
Solving the equations for $m < 0$ is instead much easier and no
particular care is needed.  For the sake of simplicity we have used
the same algorithm as for $m \ge 0$ (which now produces many
repetitions in the populations) and we have simply checked the
validity of our results by changing the number and size of
populations.

As explained in Sec.~\ref{sec_1rsb_m0} the cavity field distributions
can have a positive weight on ``hard'' fields, i.e.\ on fields that
constrain a variable to take either value $+1$ or $-1$ in all
solutions of the cluster.  This corresponds to $\varphi=0$, or $\psi
\in \{0,1\}$.  This would show up into a positive fraction of the
sample taking value, say, $\varphi=0$, thus leading to an inefficient
representation.

In order to circumvent this problem, we kept track explicitly of the
weights on $\varphi=0$ and $\psi \in \{0,1\}$, in analogy with
Eqs.~(\ref{eq_1RSB_Q_T0}), (\ref{eq_1RSB_P_T0}).  This also allows to
locate more precisely the appearance of a positive fraction of hard
fields in the distributions, as discussed in
Sec.~\ref{sec_frozenvar}. There is unfortunately one drawback to this
approach.  Consider Eq.~(\ref{eq:hP}) and suppose that all the
distributions $\hQ$ of the right hand sides are supported on
$\varphi_i^\pm\in (0,1]$. By definition the fields $\psi$ thus
generated are also strictly positive. However the degrees $l_{\pm}$
are of order $\alpha k$ (i.e.\ around $40$ for $4$-SAT in the 1RSB
regime).  As a consequence, it may happen that the product of the
$l_-$ fields $\varphi^-_i$ is smaller than the smallest number in the
computer representation used (using 64 bits and the denormalized
floating point notation this limit is roughly $\psi_\text{min} \approx
5\cdot 10^{-324}$).  How should one treat such cases?  We have adopted
the solution of ignoring, that is not including it in the population,
any number below $\psi_\text{min}$.  This solution is equivalent to
saying that we are describing with a finite population of numbers the
distribution $\hP(\psi)$ not on the domain $\psi \in (0,1]$, but on
the domain $\psi \in (\psi_\text{min},1]$.

A different solution could be to convert all the numbers smaller than
$\psi_\text{min}$ to zero.  We have tried this procedure, but it 
seems to be unstable, and to introduce systematic errors.
In particular one obtains a positive weight for $\psi=0$, even for
values of the parameters for which this is inconsistent.

The last point we would like to discuss is the problem of how to
initialize the population dynamics algorithm. It is clear that an
iterative procedure does in general lead to different solutions
depending on the starting point of the iterations. For instance the RS
solution, where the distributions $P(h)$ in $\P_{(1)}$ are
concentrated on a single value of $h$, is always a fixed point of the
1RSB equations. In the case $m=1$, the interpretation in terms of tree
reconstruction~\cite{MeMo} leads to a clear prescription for this
initialization, as explained in more details in
Sec.~\ref{sec_1rsb_m1}. One can follow the same procedure for other
values of $m$, namely initialize the populations with essentially only
hard fields. This is crucial in particular for $k=3$, where softer
initial conditions lead to an unphysical fixed point,
cf. App.~\ref{appendix_k3}.

\section{Large $k$ analysis: some technical details}
\label{app_largek}

In this appendix we provide the complete formulae for the dynamical
transition regime of Section \ref{sec:DynamicalTr}.  To leading order
one can write a set of coupled equations for the average of
$\tP(\,\cdot\,)$, $\tQ(\,\cdot\,)$ over the 1RSB order parameters
$\P_{(1)}$, $\Q_{(1)}$.  With a slight abuse of notation we shall keep
denoting by $\tP$, $\tQ$ such averages.  In terms of this quantities
we have
\begin{eqnarray}
\Lambda(\delta,m) = -\log\left\{\int\!\de\tP(h)\, \left(\frac{1+\tanh
      h}{2}\right)^m\right\} \ ,
\end{eqnarray}
where the dependence on $\delta$ is through $\tP$.  Notice that
$\Lambda(\delta,m=0)=0$ independently of $\tP$.  For $m=1$ one can use
the fact that by symmetry $\int \de\tP(h)\, (\tanh h) = 0$, to deduce
$\Lambda(\delta,m=1) =\log 2$.

For $m\neq 0,1$ one has to determine the distributions $\tP$ and
$\tQ$.  It turns out that
\begin{eqnarray}
\tQ(u) = \left(1-\frac{w}{2^k\log k}\right)\,\delta(u) + 
\frac{w}{2^k\log k}\, \tQ'(u)\, ,\qquad w\equiv 2^{2-m}\delta
\int \!\de \tP(h) \left(\frac{3+\tanh h}{2}\right)^m\, .
\end{eqnarray}
The distributions $\tP$, $\tQ'$ are solutions of the coupled equations
\begin{eqnarray}
\tQ'(u) & = & \frac{2^{2-m} \delta }{w}\int \de\tP(h) \,
 \left(\frac{3+\tanh h}{2}\right)^m\; \delta\left(
u-\frac{1}{2}\log(1+e^{-2h})\right)\, , \\
\tP(h) & = &\frac{1}{Z}\E_{l_{\pm}}
\int\!\prod_{i=1}^{l_+}\de\tQ'(u_i^+)\int\!\prod_{i=1}^{l_-}\de\tQ'(u_i^-)
\,
z_3(u_1^+,\dots,u_{l_+}^+,u_1^-,\dots,u_{l_-}^-)^m
\;
\delta\left(h-\sum_{i=1}^{l_+} u_i^+ + \sum_{i=1}^{l_-} u_i^-\right)\, ,
\end{eqnarray}
where in the second equation $Z$ is a normalizing factor and $l_\pm$
are two independent Poisson random variables of mean $w/2$.

\section{Non-uniqueness of solutions of the 1RSB equations for $k=3$}
\label{appendix_k3}

\begin{figure}
\includegraphics[width=0.6\textwidth]{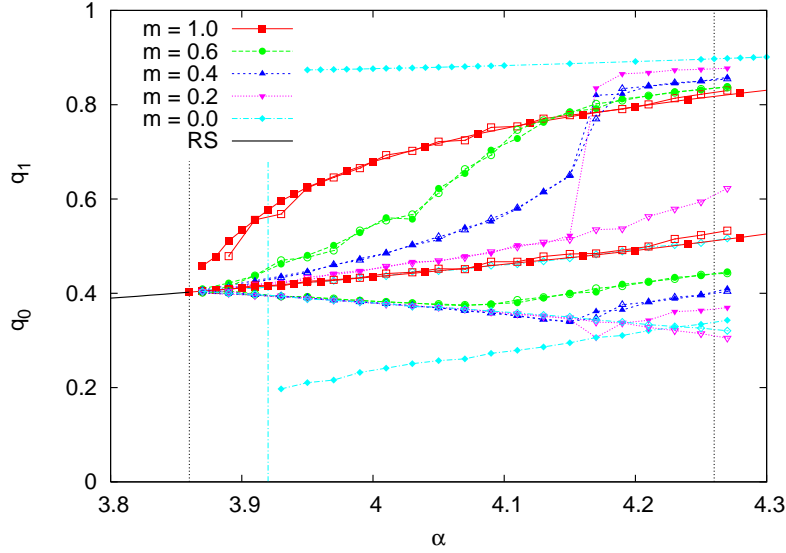}
\caption{Intra and inter-states overlap, $q_0$ and $q_1$, for $k=3$
  and some values of the Parisi parameter $m$.  Data below (resp.\ above) the RS line
  are for $q_0$ (resp.\ $q_1$).  Full (resp.\ open) symbols refer to data measured
  while increasing (resp.\ decreasing) $\alpha$.}
\label{fig:k3_q0q1}
\end{figure}

This appendix provides further details on the numerical solution of
the 1RSB equations for $k=3$. A difficulty that arises in this case is
the presence, for some values of $\alpha$ and $m$, of at least two
distinct non-trivial solutions of the 1RSB equations (this has been
already noticed in~\cite{k3_Zhou} for $\alpha=4.2$, and in~\cite{FlLe}
for the related coloring problem). As a consequence the initial
conditions of the iterative resolution play an important role in
selecting the fixed point that shall be reached.

One can justify the existence of multiple solutions as follows. 
As mentioned in the main text, the continuous dynamical transition at
$\alpha_{\rm d}\approx 3.86$ corresponds to a local instability of the
RS solution with respect to 1RSB perturbations.  It is important to
underline that this instability condition is independent on the value
of $m$, that is at $\alpha_{\rm d}$ a new solution of the 1RSB
equations should grow continuously away from the RS one, for all
values of $m$.  This is illustrated in Fig.~\ref{fig:k3_q0q1}, where
the overlaps $q_0$ and $q_1$ meet at $\alpha_{\rm d}$ for various
values of $m$.  By continuity these solutions do not contain hard
fields in the neighborhood of $\alpha_{\rm d}$. On the contrary it is
known since~\cite{MeZe} that another solution of the $m=0$ equations,
with a finite weight on hard fields, arises discontinuously at $\alpha
\approx 3.92$. For larger values of $\alpha$ these two solutions thus
coexist\footnote{Let us signal a peculiarity of the $m=0$ `soft'
  solution.  It is easy to realize from
  Eqs.~(\ref{eq_1RSB_Q_T0},\ref{eq_1RSB_P_T0}) that the average of the
  distributions $P(h)$ and $Q(u)$ in this solution verify the RS
  equations. In consequence its intra-overlap $q_1$ coincides with the
  RS overlap, its complexity vanishes and its internal entropy equals the 
  RS one.}.  A natural conjecture is that two solutions also
coexist for $m \neq 0$.  The iterative population dynamics algorithm
converges to one of them depending on the initialization (more
precisely, on the fraction of hard fields in the initial populations).

Our data suggest that the interval of $\alpha$ in which the two
solution coexist shrinks when $m$ grows from $0$.  For instance in
Fig.~\ref{fig:k3_q0q1} one clearly see two branches for $m=0.2$ at
high enough values of $\alpha$, whereas for $m=0.6$ the two curves
obtained by increasing and decreasing $\alpha$ at fixed $m$ are
superimposed within numerical precision.

\begin{figure}
\includegraphics[width=0.6\textwidth]{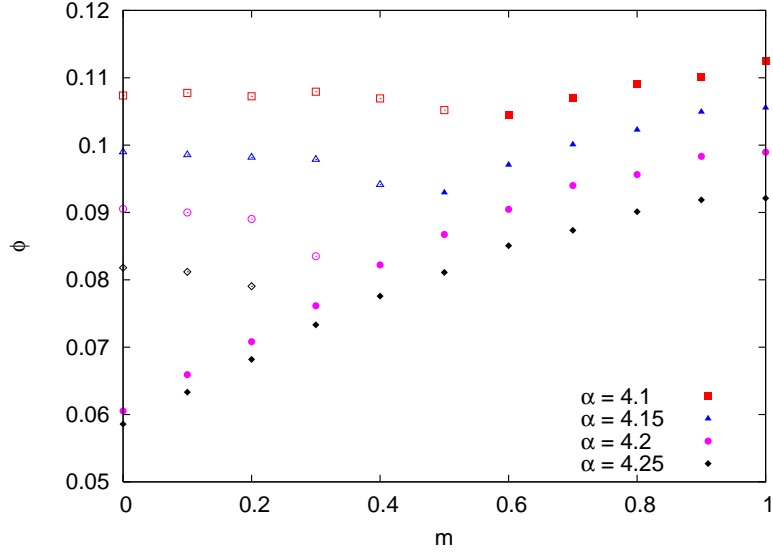}
\caption{The internal entropy should be a non-decreasing function
  of $m$ if the solution is consistent. Filled (resp.\ empty) symbols
  refer to solutions with $\partial_m \phi>0$
  (resp.\ $\partial_m \phi<0$), for $k=3$.}
\label{fig:k3-phi}
\end{figure}

\begin{figure}
\includegraphics[width=0.6\textwidth]{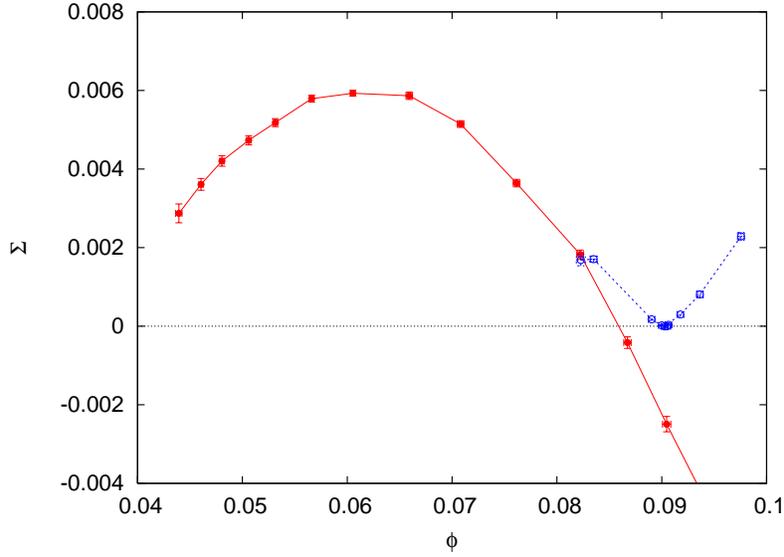}
\caption{The entropic complexity $\Sigma(\phi)$ for $k=3$ and
  $\alpha=4.2$.  The two different branches correspond to the
  consistent (full line) and inconsistent solution (dashed line).}
\label{fig:k3-Sigma_phi_a42}
\end{figure}

It remains to understand which, if any, of these solutions is the
correct one. In principle one should test their stability with respect
to higher level of replica symmetry breaking~\cite{MoPaRi,MoRi,FlLe2},
however it is an extremely demanding numerical task that we did not
undertake. A simpler consistency argument can be invoked by computing
the internal entropy of the pure states.  This should be an increasing
function of $m$.  We can see on the curves of Fig.~\ref{fig:k3-phi}
that this condition is not respected for all the values of $\alpha$
and $m$ (full symbols refer to consistent solutions, while open symbol
are for inconsistent ones).  For values of $\alpha$ smaller than
roughly $4.15$ we are not able to find a consistent solution in the
whole range of $m\in[0,1]$ (a consistent solution exists only for $m$
large enough).  While for $\alpha$ roughly larger than $4.15$ two
solutions coexist at small values of $m$ and the consistent one is the
one with more hard fields.  We also notice that this inconsistency is
accompanied by the decreasing of the inter-overlap $q_0$ with
$\alpha$: in other words we empirically find that the quantities
$\partial_m \phi$ and $\partial_\alpha q_0$ always have the same sign.
This observation makes easier to locate in Fig.~\ref{fig:k3_q0q1}
consistent solutions (those with $q_0$ increasing with $\alpha$).  In
order to make connection with previous studies where consistent and
inconsistent solutions were found~\cite{BiMoWe,MePa_T0,FlLe} we plot
in Fig.~\ref{fig:k3-Sigma_phi_a42} the entropic complexity curve for
$\alpha=4.2$: the full (resp. dashed) curve corresponds to the
consistent (resp. inconsistent) branch.


\end{document}